\documentclass[showpacs,amsmath,amssymb,twocolumn,prb]{revtex4}
\usepackage{graphicx}
\usepackage{amsfonts}
\usepackage{amsmath, amsthm, amssymb, bbm}
\usepackage{times}
\usepackage{subfigure}

\newcommand{\ve}[1]{\boldsymbol{#1}}
\DeclareMathOperator{\diag}{diag} 

\newcommand{\e}[1]{\mathrm{e}^{#1}}

\newcommand{\bq}{\begin{equation}}
\newcommand{\eq}{\end{equation}}
\newcommand{\eps}{\varepsilon}

\newcommand{\g}{\underline{\gamma}}
\newcommand{\gt}{\underline{\tilde{\gamma}}}
\newcommand{\N}{\underline{\mathcal{N}}}
\newcommand{\Nt}{\underline{\tilde{\mathcal{N}}}}

\newcommand{\Smat}{\hat{\mathcal{S}}}
\newcommand{\ppara}{\boldsymbol{p}_\parallel}

\newcommand{\eg}{\textit{e.g. }}

\def\i{\mathrm{i}}

\begin{document}
\title[
Signature of odd-frequency pairing correlations induced
by a magnetic interface
]{
Signature of odd-frequency pairing correlations induced
by a magnetic interface
}
\author{Jacob Linder}
\affiliation{Department of Physics, Norwegian University of
Science and Technology, N-7491 Trondheim, Norway}
\author{Asle Sudb{\o}}
\affiliation{Department of Physics, Norwegian University of
Science and Technology, N-7491 Trondheim, Norway}
\author{Takehito Yokoyama} 
\affiliation{Department of Physics, Tokyo Institute of Technology, 2-12-1 Ookayama, Meguro-ku, Tokyo 152-8551, Japan}
\author{Roland Grein$^{(1)}$}
\affiliation{
$^{(1)}$Institut f{\"u}r Theoretische Festk{\"o}rperphysik and DFG-Center for Functional Nanostructures,
Karlsruhe Institute of Technology, D-76128 Karlsruhe, Germany\\
$^{(2)}$Fachbereich Physik, Universit\"at Konstanz D-78457 Konstanz, Germany}
\author{Matthias Eschrig$^{(1,2)}$}
\affiliation{
$^{(1)}$Institut f{\"u}r Theoretische Festk{\"o}rperphysik and DFG-Center for Functional Nanostructures,
Karlsruhe Institute of Technology, D-76128 Karlsruhe, Germany\\
$^{(2)}$Fachbereich Physik, Universit\"at Konstanz D-78457 Konstanz, Germany}

\date{Received \today}
\begin{abstract}
\noindent We investigate the mutual proximity effect in a normal metal contacted to a superconductor through a magnetic interface. 
Analytical and self-consistent numerical results are presented, and we consider both the diffusive and ballistic regimes. 
We focus on the density of states in both the normal and superconducting region, and find that the presence of spin-dependent 
phase-shifts occurring at the interface qualitatively modifies the density of states. In particular, we find that the 
proximity-induced pairing amplitudes in the normal metal region undergo a conversion at the Fermi level from pure even-frequency 
to odd-frequency. Above a critical value of the interface spin-polarization (or, equivalently, for fixed interface spin-polarization, 
above a critical interface resistance), only odd frequency correlations remain. This is accompanied by the replacement of the 
familiar proximity minigap or pseudogap in the normal layer by an enhancement of the density of states above its normal state 
value for energies near the chemical potential. The robustness of this effect towards inelastic scattering, impurity scattering, 
and the depletion of the superconducting order parameter close to the interface is investigated. We also study the inverse 
proximity effect in the diffusive limit. We find that the above-mentioned conversion persists also for thin superconducting layers 
comparable in size to the superconducting coherence length $\xi_\text{S}$, as long as the inverse proximity effect is relatively 
weak. Concomitantly, we find a shift in the critical interface resistance where the pairing conversion occurs. Our findings suggest a 
robust and simple method for producing purely odd-frequency superconducting correlations, that can be tested experimentally.
\end{abstract}

\maketitle

\section{Introduction}

The proximity effect in hybrid structures with superconductors offers an arena of interesting physics to explore which could also prove 
to be useful in nanotechnological devices. The incorporation of ferromagnetic elements in such hybrid structures activates the spin degree 
of freedom, which has a number of important consequences for how the proximity effect is manifested in physical quantities.\cite{bergeretrmp,buzdinrmp,eschrig07} 
In the case of a ferromagnet$\mid$superconductor (F$\mid$S) bilayer, it 
is known that so-called odd-frequency pairing is generated. \cite{bergeretPRL} 
Odd-frequency pairing has been studied previously 
\cite{berezinskii,bal92,abr95,coleman93,fuseya03} 
in particular in connection with the search for exotic superconducting states 
that may arise via the mechanism of spontaneous symmetry breaking. 
A particular feature of such odd-frequency paring states is
a strong retardation effect which makes the equal-time correlator vanish for the 
Cooper pair. 

Apart from the mechanism of spontaneous symmetry breaking, 
odd-frequency pairing correlations can also be 
created by an {\it induced} symmetry breaking.
The general requirement for such a generation of odd-frequency correlations is that either translational symmetry 
(for odd-frequency singlet),\cite{eschrig07,tanakaPRL,Yokoyama,Yokoyama2} or both translational and spin-rotational symmetry 
(for odd-frequency triplet),\cite{bergeretPRL,eschrig07,yokoyama07,eschrig08,linderyokoyama_prb_08,linderzareyan_prb_09} 
are explicitely broken. As a result, 
one would expect to see odd-frequency superconductivity as a quite generic feature of proximity structures.  Although this 
fact is well known since long among the community dealing with inhomogeneous problems in superconductivity, it is only recently 
that the attention has shifted to the question: how may one extract and detect these exotic pairing correlations, and in particular 
the odd-frequency triplet state,  experimentally?
\par
There are two major difficulties associated with the detection of the odd-frequency triplet state. One obstacle is that such a state 
induced in F$\mid$S bilayers often has a very short penetration depth into the ferromagnetic region of order $\sim \mathcal{O}$(nm). 
In fact, unless there are magnetic inhomogeneities present in the interface region,\cite{eschrig03} it is limited by the magnetic 
coherence length $\xi_{\rm F}$ which usually is much smaller than the superconducting coherence length $\xi_\text{S}$. A second obstacle 
related to the detection of odd-frequency correlations is that these often compete with even-frequency superconducting correlations in 
the same material, masking their presence. To find smoking gun signatures of odd-frequency pairing is therefore a rather challenging 
issue to tackle, although there are a few experimental works which have pointed towards fingerprints of odd-frequency pairing.\cite{keizer_nature_06,sosnin06}
\par
Recently, it has been realized that the interface properties in hybrid structures with superconductors play a pivotal role in magnetic 
aspects of the proximity effect.\cite{greinPRL,greinPRB} In most works, non-magnetic (or spin-inactive) interfaces have been considered, 
even in the presence of ferromagnetic elements. Utilizing the quasiclassical theory of superconductivity, such interfaces are modeled as 
effective boundary conditions. For the general Eilenberger equation, boundary conditions for non-magnetic systems and spin-inactive 
interfaces were first formulated in implicit form in Refs.~\onlinecite{zaitsev84} and \onlinecite{shelankov84}. An explicit formulation 
has been derived in Ref.~\onlinecite{eschrig00}. For the diffusive limit of the theory, described by the Usadel equation, boundary 
conditions have been formulated in Refs.~\onlinecite{kupluk} and \onlinecite{nazarov99}.

However, the spin-dependent properties of the interface may become important when ferromagnetic elements are present in 
the system. In particular, the transmission properties of spin-$\uparrow$ and spin-$\downarrow$ electrons into a 
ferromagnetic metal are different, which gives rise to both spin-dependent conductivities (spin filtering) \cite{meservey} 
and spin-dependent phase shifts (spin-DIPS) at the interface.\cite{Tokuyasu88,fogelstrom00,hh,eschrig03,zhao04,audrey,bobkova07,linder_prb_07,eschrig08,brydon,lindercuoco_arxiv_10}
A generalization of boundary conditions to spin-active interfaces
was given in Refs.~\onlinecite{Millis,fogelstrom00,zhao04}, that has been 
generalized to include systems with strong exchange splitting of the
energy bands in Refs.~\onlinecite{eschrig03} and \onlinecite{eschrig09}.
\par
The spin-DIPS can lead to qualitatively novel effects in superconducting hybrid systems. Very recently, the proximity effect in a 
normal metal$\mid$superconductor (N$\mid$S) bilayer with a magnetic interface was studied in Ref.~\onlinecite{linder_prl_09}, and 
a surprising result was unveiled. Namely, above a critical interface resistance, the proximity-induced superconducting correlations 
in the normal metal at the Fermi level change abruptly from conventional even-frequency pairing to odd-frequency pairing. This 
result is interesting for two reasons. Firstly, the odd-frequency correlations penetrate much deeper into the normal metal region, 
since there is no explicit exchange field there. Secondly, the result provides a scenario where odd-frequency amplitudes are present 
without any interfering effects of even-frequency correlations. In light of the above discussion, it is seen that this actually 
resolves the two main difficulties associated with the experimental detection of odd-frequency correlations.

\begin{figure}[t!]
\centering
\resizebox{0.45\textwidth}{!}{
\includegraphics{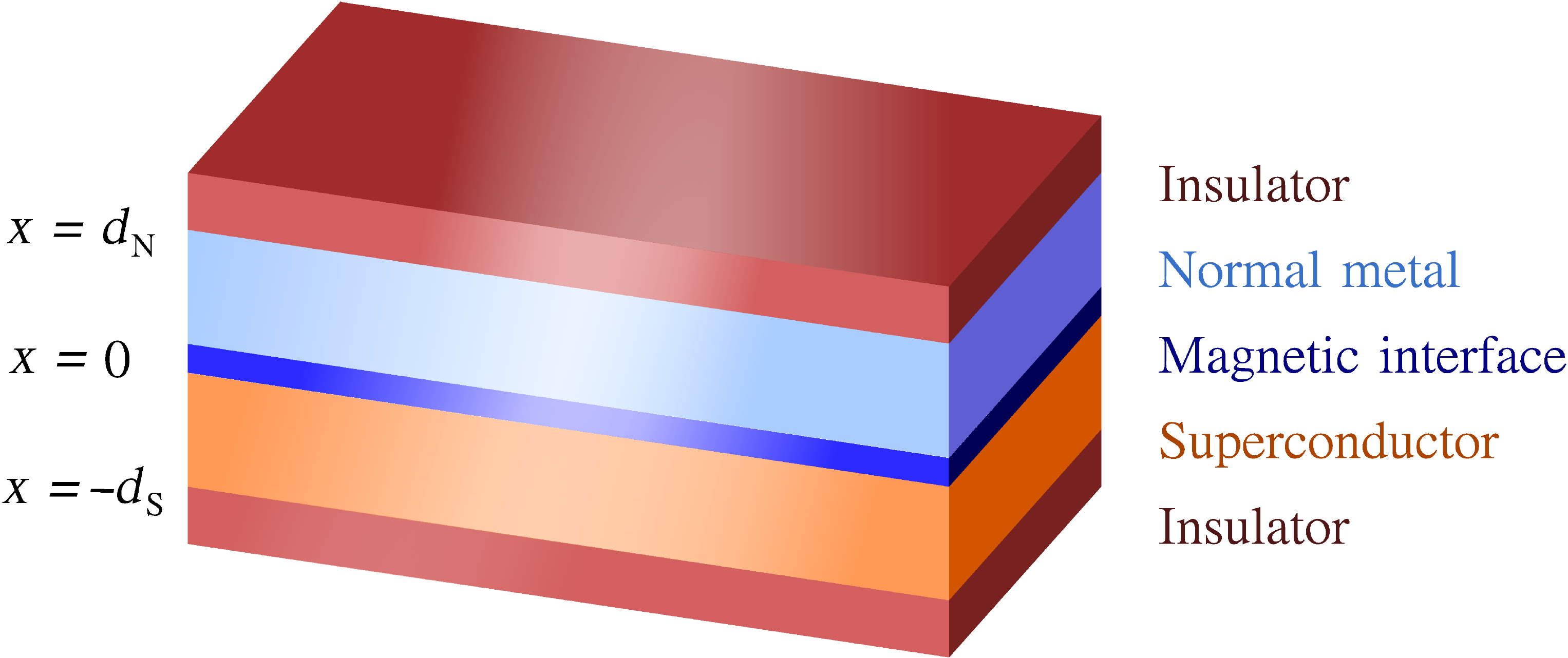}}
\caption{(Color online) Proposed experimental setup for observation of the odd-frequency component in a normal metal layer$\mid$superconductor junction.}
\label{fig:model} 
\end{figure}

\par
In this work, we expand on the results provided in Ref.~\onlinecite{linder_prl_09} and address in particular three complementary issues: \textit{(i)} how is the even- to odd-frequency conversion influenced by pair breaking effects near the interface, \textit{(ii)} how is the inverse proximity effect in the superconducting region influenced by the presence of spin-DIPS, and
\textit{(iii)} how does a Fermi surface mismatch influence the effect under consideration? 
These questions are important from an experimental perspective, where non-idealities such as pair breaking effects are generically present, and demand a numerical and self-consistent approach. The system under consideration is shown in Fig. \ref{fig:model}. The superconductor is assumed to be a conventional superconductor such as Al or Nb, thus featuring a spin-singlet, even-parity (isotropic), even-frequency symmetry for the order parameter. The insulating interface region separating the normal metal and the superconductor is assumed to be magnetic, \eg EuO. The density of states (DOS) can be probed experimentally in various ways, for instance spectroscopically 
by using a local scanning tunneling microscopy (STM)-tip.
\par
This work is organized as follows. In Sec. \ref{sec:theory_diff} and \ref{sec:theory_ball}, we establish the theoretical framework to be 
used for obtaining our results. 
In Sec. \ref{sec:results_diff} and \ref{sec:results_ball}, we present our main results, demonstrating that the even-odd 
frequency conversion is a robust effect which survives both in the clean and dirty limit, and moreover is resilient towards pair-breaking 
effects near the interface. We summarize our findings in Sec. \ref{sec:summary}. 
We shall use units such that $\hbar$=$c$=$k_\text{B}$=1. Moreover, we use $\underline{\bullet }$ for 2$\times$2 spin-matrices, 
$\hat{\bullet }$ for 4$\times$4 matrices in Nambu-Gor'kov particle-hole space, and boldface notation for vectors.

We use the quasiclassical theory of superconductivity,\cite{qcl1,qcl2,qcl3,qcl4,qcl5} where information about the physical properties 
of the system is embedded in the Green's function. For an equilibrium situation, it suffices to consider the retarded part of the 
Green's function, here denoted $\hat{g}$. We begin our discussion with the diffusive limit, after which we proceed to the ballistic 
case. 

\section{Diffusive limit}\label{sec:diffusive}

\subsection{Theory}\label{sec:theory_diff}
Due to the symmetry properties of $\hat{g}$, one may parameterize it conveniently in the superconducting (S) and normal (N) region 
in the diffusive limit.\cite{linder_prb_08}  Consider for concreteness an N$\mid$S bilayer, where we may write
\begin{align}\label{eq:gS}
\hat{g}_\text{S} = \begin{pmatrix}
c & 0 & 0 & s \\
0 & c & -s & 0\\
0& s & -c & 0\\
-s & 0 & 0 & -c \\
\end{pmatrix}
\end{align} 
with $c=\cosh(\theta)$, $s=\sinh(\theta)$, and $\theta=\text{arctanh}(\Delta/\varepsilon)$.
In the normal region one finds
\begin{align}\label{eq:gN}
\hat{g}_\text{N} = \begin{pmatrix}
c_\uparrow & 0 & 0 & s_\uparrow\\
0 & c_\downarrow & s_\downarrow & 0\\
0& -s_\downarrow & -c_\downarrow & 0\\
-s_\uparrow & 0 & 0 & -c_\uparrow\\
\end{pmatrix},
\end{align} 
with $c_\sigma=\cosh(\theta_\sigma)$, $s_\sigma=\sinh(\theta_\sigma)$. The diffusive propagators are normalized according to
$\hat g_\text{S}^2= \hat g_\text{N}^2=\hat{1}$ where $\hat{1}=\diag(1,1,1,1)$.
Through this parameterization, we have taken into account the possibility of odd-frequency triplet correlations in the normal region, while 
we have employed the bulk solution in the superconductor. This approximation is valid under the assumption that the superconducting layer 
is much thicker and less disordered than the normal region, thus acting as a reservoir.\cite{bergeretrmp} The gap suppression near the 
interface may furthermore be neglected in the tunneling limit.\cite{bruder} In general, the superconducting region is also influenced 
by the proximity effect, in which case a similar parameterization as Eq. (\ref{eq:gN}) is employed also in that region. We will return 
to this issue below.

In the present case, the Green's function $\hat{g}_\text{N} $
in the normal region obeys the Usadel equation
\begin{align}\label{eq:usadel_original}
D\nabla(\hat{g}_\text{N}\nabla\hat{g}_\text{N}) + \i[\varepsilon\hat{\rho}_3,\hat{g}_\text{N}] = 0,
\end{align}
with $\hat{\rho}_3=\diag(1,1,-1,-1)$,
and is subject to boundary conditions at the S$\mid$N $(x=0)$ and N$\mid$I $(x=d_\text{N})$ interfaces as follows:\cite{hh,audrey}
\begin{align}\label{eq:bc}
2\gamma d \hat{g}_\text{N} \partial_x \hat{g}_\text{N} = [\hat{g}_\text{S},\hat{g}_\text{N}] + \i \frac{G_\phi}{G_T} [\hat{\tau}_3, \hat{g}_\text{N}]
\end{align}
with $\hat{\tau}_3=\diag(1,-1,1,-1)$, at $x=0$ and $\partial_x\theta_\sigma=0$ at $x=d_\text{N}$. Here, $\gamma = R_\text{B}/R_\text{N}$ where $R_\text{B}$ $(R_\text{N})$ is the resistance of the barrier (normal region), and $d_\text{N}$ is the width of the normal region, while $G_T$ 
is the barrier conductance. For later use, we define the superconducting coherence length $\xi_\text{S} = \sqrt{D/\Delta}$ and Thouless energy $\varepsilon_\text{Th}=D/d_\text{N}^2$, where $D$ is the diffusion constant. Eq. (\ref{eq:bc}) contains an additional term $G_\phi$ compared to the usual non-magnetic boundary conditions in Refs.~\onlinecite{kupluk} and \onlinecite{nazarov99}. The physical interpretation of this term is that it gives rise to spin-dependent phase shifts of quasiparticles being reflected at the interface. Note that $G_\phi$ may be non-zero even if the transmission $G_T\to0$, corresponding to a ferromagnetic insulator.\cite{hh}  Later in this work, we shall also consider a fully self-consistent calculation where the bulk solution is \textit{not assumed} in the superconducting region.
\par
\begin{figure}
\centering
\resizebox{0.49\textwidth}{!}{
\includegraphics{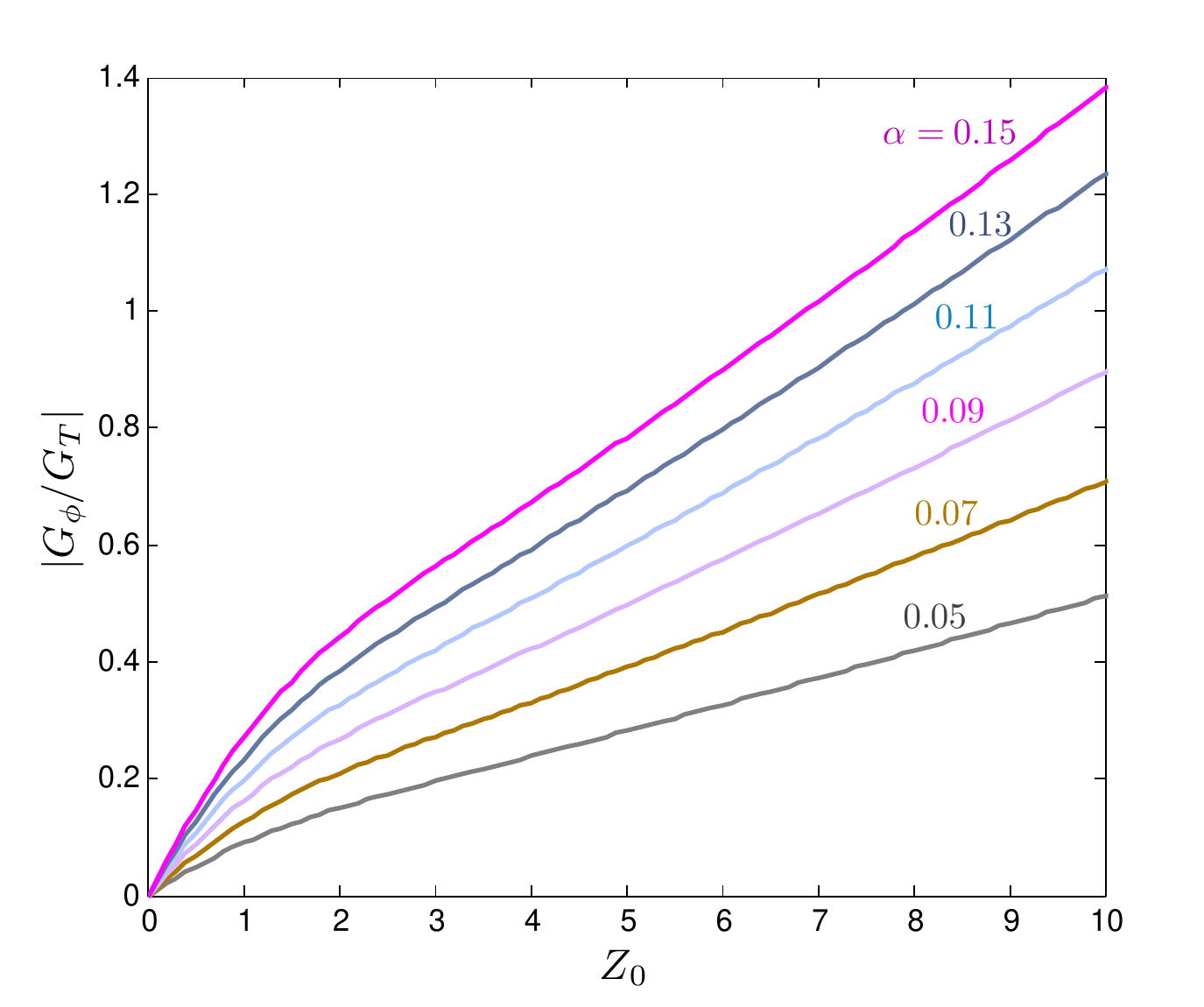}}
\caption{(Color online) Plot of the ratio $|G_\phi/G_T|$ as a function of the barrier strength $Z_0$ for several values of the interface polarization $\alpha$.}
\label{fig:Gphi} 
\end{figure}

Using a simplified scattering model near the interface, it is possible to obtain microscopic expressions for $G_T$ and $G_\phi$.
They are related to the transmission and reflection amplitudes $\{t_{\sigma}^\text{\text{S}(\text{N})}, r_{\sigma}^\text{\text{S}(\text{N})}\}$ 
on the S (N) side of the interface. For simplicity, we assume that the interface is characterized by $N$ identical scattering channels. 
Under the assumption of tunnel contacts, one obtains from a model with a Dirac-like barrier potential 
 \begin{align}
 G_T = NG_QT,\; G_\phi = 2NG_Q\Big( \rho^\text{N} - 4\tau^\text{S}/T\Big)
 \end{align}
 upon defining $T = \sum_\sigma |t_{\sigma}^\text{S}|^2$, $G_Q = e^2/(2\pi\hbar)$, and 
 \begin{align}
 \rho^\text{N} = \text{Im}\{r_{\uparrow}^\text{N} (r_{\downarrow}^\text{N})^*\},\; \tau^\text{S} = \text{Im}\{t_{\uparrow}^\text{S} (t_{\downarrow}^\text{S})^*\}.
 \end{align}
The scattering coefficients take the form:
 \begin{align}
 r_\sigma^\text{N} &= (k^\text{N} - k^\text{S} - \i k^\text{S}Z_\sigma)/\mathcal{D}_\sigma,\; t_\sigma^\text{S} &= 2\sqrt{k^\text{S} k^\text{N}}/\mathcal{D}_\sigma,
 \end{align}
 with the definitions $\mathcal{D}_\sigma = k^\text{S} + k^\text{N} + \i k^\text{S}Z_\sigma$ , $k^\text{S} = \sqrt{2m_\text{S}\mu_\text{S}}$, $k^\text{N} = \sqrt{2m_\text{N}\mu_\text{N}}$. 
 Here, $Z_\sigma = Z_0 + \sigma Z_\text{s}$ is the spin-dependent barrier potential, and we define  $\alpha = Z_\text{s}/Z_0$ as the 
 polarization for the barrier. The ratio $|G_\phi/G_T|$ is evaluated in Fig. \ref{fig:Gphi} as a function of the barrier strength $Z_0$ 
 for several values of $\alpha$. We have used $\mu_\text{S}=\mu_\text{N}$ = 5 eV and set $m_\text{S}=m_\text{N}$ to the bare electron 
 mass. A Fermi-vector mismatch $\mu_\text{S}\neq\mu_\text{N}$ between the materials is accounted for by an increase in $Z_0$. As seen, 
 the ratio $|G_\phi/G_T|$ can be of order unity for low barrier transparencies $Z_0\gg1$ even for relatively weak polarizations with 
 $\alpha=10\%$.

\subsection{Results}\label{sec:results_diff}

We begin our analysis by discussing the weak proximity regime, where an analytical treatment is possible for all quasiparticle energies $\varepsilon$. Thereafter, we present a self-consistent numerical calculation for an arbitrary proximity effect, incorporating pair-breaking mechanisms and the depletion of the superconducting order parameter near the interface region. In the linearized treatment, one assumes 
that the deviation from the bulk Green's function in the ferromagnetic region is small. This permits us to write the retarded Green's 
function on the form
\begin{align}
\hat{g}^R \simeq \hat{g}_0 + \hat{f},\; \hat{g}_0 = \hat{\rho}_3.
\end{align}
Here, we have defined
\begin{align}
\hat{f} &= \begin{pmatrix} 
\underline{0} & \underline{f}(\varepsilon) \\
-[\underline{f}(-\varepsilon)]^* & \underline{0}\\
\end{pmatrix},
\notag\\
 \underline{f}(\varepsilon) &= 
\begin{pmatrix}
0 & f_+(\varepsilon) \\
f_-(\varepsilon) & 0\\
\end{pmatrix}.
\end{align}
Under the assumption of an equilibrium situation, the Keldysh Green's function is given by
\begin{align}
\hat{g}^K = [\hat{g}^R - \hat{g}^A ]\tanh(\beta\varepsilon/2),
\end{align}
where $\beta=1/T$ is inverse temperature. The advanced component is $\hat{g}^A = -(\hat{\rho}_3\hat{g}^R\hat{\rho}_3)^\dag$. The linearized Usadel equation \cite{usadel} may be written as
\begin{align}
D\partial_x^2 f_\pm + 2\i\varepsilon f_\pm = 0,
\end{align}
and is to be supplemented with the boundary condition obtained from Eq. (\ref{eq:bc}) 
\begin{align}
\gamma d_\text{N}\partial_x f_\pm = (cf_\pm \mp s) \pm \i \frac{G_\phi}{G_T} f_\pm
\end{align}
at $x=0$ while $\partial_x f_\pm = 0$ at $x=d_\text{N}$. Here, $f_\pm = f_{\rm t}\pm f_{\rm s}$ where $f_{\rm t}$ is the $S_z=0$ triplet component and $f_{\rm s}$ is the singlet component of the anomalous Green's function. Since the diffusive limit is considered, the singlet component has an even-frequency symmetry while the triplet component has an odd-frequency symmetry. The odd-frequency component has previously been predicted to appear in S$\mid$F layers, but we now show that the presence of a magnetically active barrier region induces an odd-frequency component in a S$\mid$N layer, with a much longer penetration depth. We find that the solution for the Green's function reads 
\begin{align}\label{eq:lin}
f_\pm &= \frac{\pm s[\e{\i k(x-2d_\text{N})} + \e{-\i kx}]}{\i k\gamma d_\text{N} (1 - \e{-2\i kd_\text{N}}) + (c \pm \i G_\phi/G_T)(1 + \e{-2\i kd_\text{N}})}.
\end{align}
Here, $k=\sqrt{2\i\varepsilon/D}$. For a spin-inactive barrier, $G_\phi = 0$, we obtain $f_+ = -f_-$, such that $f_{\rm t}=0$. However, the presence of $G_\phi$ induces the odd-frequency component in the normal layer. The decay length here is not dictated by the magnetic coherence length $\xi_\text{F} = \sqrt{D/h}$ as in an S$\mid$F layer, but by $\xi_\text{N} = \sqrt{D/\varepsilon }$ as in an
S$\mid$N layer.
This allows the odd-frequency component to penetrate much deeper into the N layer than into the F layer. The simplest experimental manifestation of the odd-frequency component is probably a zero-energy peak in the local density of states.\cite{Asano,yokoyama07,Braude} In S$\mid$F layers, where this phenomenon has been discussed previously, a clear zero-energy peak is unfortunately often masked by the simultaneous presence of singlet correlations $(f_{\rm s})$, which tend to suppress the density of states at low energies. In the present case of a spin-active interface in an S$\mid$N junction, however, Eq. (\ref{eq:lin}) suggests a remarkable effect. Consider $\varepsilon=0$, for which $k=0$, $s=\i$, and $c=0$, leading to the result
\begin{align}
f_\pm = G_T/G_\phi
\end{align}
under the assumption that $G_\phi \neq 0$. This equation conveys a powerful message, namely that \textit{at the Fermi level, 
the singlet component is absent while the triplet component remains. Moreover, the latter is determined simply by the ratio of $G_T$ 
and $G_\phi$}. 
Consequently, this should provide ideal circumstances for direct observation of the odd-frequency component, manifested as a zero-energy 
peak in the local density of states.

So far, we have limited ourselves to the weak proximity effect regime. We now consider an arbitrarily large proximity effect. In 
this case, the Usadel equation reads 
\begin{align}
D\partial_x^2\theta_\sigma + 2\i\varepsilon\sinh\theta_\sigma= 0,
\end{align}
while the boundary conditions become 
\begin{align}
\gamma d_\text{N} \partial_x\theta_\sigma = (c s_\sigma - \sigma sc_\sigma) + \i\sigma \frac{G_\phi}{G_T} s_\sigma
\end{align}
at $x=0$ and $\partial_x\theta_\sigma = 0$ at $x=d_\text{N}$. A general analytical solution of the above equation can hardly be 
obtained, but it may be solved at zero energy. For $\varepsilon =0$ we find pairing amplitudes that are either purely 
(odd-frequency) triplet for $|G_\phi|>G_T$,
\begin{align}\label{eq:full1}
\textstyle
f_{\rm s}(0) = 0,\quad f_{\rm t}(0) =\frac{\displaystyle G_T \cdot \text{sgn}(G_\phi )}{\displaystyle \sqrt{ G_\phi \!^{\!\! 2}-G_T \,^{\! 2}}},
\end{align}
or purely (even-frequency) singlet for $|G_\phi|<G_T$,
\begin{align}\label{eq:full2}
\textstyle
f_{\rm s}(0) = \frac{\displaystyle \i\cdot G_{ T}}{\displaystyle \sqrt{G_T \,^{\! 2}-G_\phi \!^{\!\! 2}}},\quad f_{\rm t}(0) = 0.
\end{align}
Thus, the presence of $G_\phi$ induces an odd-frequency component in the normal layer. 
The remarkable aspect of Eqs. (\ref{eq:full1}) and (\ref{eq:full2}) is that they 
are valid for any value of the width $d_\text{N}$ below the inelastic scattering length, and for
any interface parameter $\gamma$. Thus, the vanishing of the singlet component 
is a robust feature in S$\mid$N structures with spin-active interfaces, as long 
as $|G_\phi|>G_T$. Without loss of generality, we focus on positive 
values of $G_\phi$ from now on. 
The DOS is given as 
\begin{align}
N(\varepsilon)/N_0 &= \sum_\sigma \text{Re}\{c_\sigma\}/2,
\end{align}
thus yielding
\begin{align}\label{eq:dosgphi}
\frac{N(\varepsilon=0)}{N_0} = \text{Re}\left\{
\frac{G_\phi}{\sqrt{G_\phi \!^{\!\! 2} -G_T \,^{\! 2}}}
\right\}.
\end{align}
At zero-energy, the DOS vanishes 
when $G_\phi<G_T $, which means that the usual 
minigap in S$\mid$N structures survives.
However, the zero-energy DOS is 
enhanced for $G_\phi>G_T $ since the singlet component vanishes there. 
\par
We suggest the following qualitative explanation for the mechanism behind the 
conversion between even- and odd-frequency correlations. The superconductor 
induces a minigap $ \propto G_T$ in the normal metal, while the spin-active 
barrier induces an effective exchange field $ \propto G_\phi$. The 
situation in the normal metal then resembles that of a thin-film conventional 
superconductor in the presence of an in-plane external magnetic field,
\cite{meservey} with the role of the gap and field played by $G_T$ and 
$G_\phi$, respectively. In that case, it is known that superconductivity is 
destroyed above the Clogston-Chandrasekhar limit,\cite{clogston} as the 
spin-singlet Cooper-pairs break up.  
In the proximity structure we consider here, Cooper-pairs persist above this 
limit as they are {\it induced} from the superconducting region where the 
exchange field is absent. However, these Cooper pairs are modified strongly by
multiple scattering from the spin-active interface, and above a
critical ratio $G_\phi/G_T=1$
spin-singlet pairing is no longer possible in the N region 
at the chemical potential. 
It is then replaced by spin-triplet pairing, which must 
be odd in frequency due to the isotropization of the correlation in the diffusive limit. 
We observe coexistence of the exchange field and spin-singlet even-frequency 
superconductivity as long as $G_\phi$ is below the critical value of $G_\phi=G_T$. 
At the critical point,
the DOS varies as $1/\sqrt{|\varepsilon |}$ and diverges at $\varepsilon =0$.
Thus, we find that 
there is a natural separation between even-frequency and odd-frequency pairing 
in the normal metal at a critical value of the effective exchange field $G_\phi$.
This agrees with the interpretation of $G_\phi$ in Ref.~\onlinecite{hh} as an effective proximity-induced exchange field.  Note that the above expressions are valid also for $G_\phi \to 0$: we obtain $f_{\rm s}=\i$ and $f_{\rm t}=0$ as demanded by consistency.
\par
The full energy-dependence of the DOS may only be obtained numerically. In addition, it is of interest to see how robust the predicted even- to odd-frequency conversion is towards the inevitable depletion of the superconducting order parameter near the interface in addition to non-ideal effects such as the presence of inelastic scattering. To investigate this, we solve the Usadel equation and the gap equation self-consistently in both the normal and superconducting region. Since we are no longer considering the bulk solution in the superconducting region, it becomes necessary to specify the width $d_\text{S}$ of the superconducting layer, the spin-dependent phase shifts $G_\phi^\text{S}$ on the superconducting side of the interface, and also the bulk resistance $R_\text{S}$ of the superconductor. The Usadel equation on the N side satisfies Eq. (\ref{eq:usadel_original}), whereas on the S side an additional term $\hat{\Delta}$ is added inside the commutator in the second term 
 of Eq. (\ref{eq:usadel_original}). Inclusion of spin-orbit coupling effects may be done similarly by including a term $\hat{\sigma}_\text{so}$ (see Ref.~\onlinecite{linder_prb_08} for a detailed treatment and expressions for such terms). The superconducting order parameter is determined self-consistently by solving the Usadel equation in conjunction with the gap equation:
\begin{align}
\Delta = \frac{N_{\rm F}\lambda}{2}\int^\omega_0\text{d}\varepsilon \tanh{(\beta\varepsilon/2)} \sum_\sigma\sigma \text{Re}\{\sinh(\theta_\sigma)\}, \end{align}
where we choose the weak coupling-constant and cut-off energy to be $N_{\rm F}\lambda=0.2$ and $\omega/\Delta_0 = 75$. Within our numerical scheme, self-consistency is typically achieved after 10 iterations. We account for inelastic scattering by the parameter $\delta/\Delta_0=10^{-3}$, where $\varepsilon\to \varepsilon+\i\delta$.

The diffusion coefficients $D_\text{N}$ and $D_\text{S}$ are in general different. At the S$\mid$N interface, the boundary condition on the normal side now reads:
\begin{align}\label{eq:bcN}
2d_\text{N}\frac{R_\text{B}}{R_\text{N}} \hat{g}_\text{N} \partial_x \hat{g}_\text{N} &= [\hat{g}_\text{S},\hat{g}_\text{N}] +\i \frac{G_\phi^\text{N}}{G_T} [\hat{\tau}_3, \hat{g}_\text{N}].
\end{align}
while on the superconducting side, one has:
\begin{align}\label{eq:bcS}
2d_\text{S}\frac{R_\text{B}}{R_\text{S}} \hat{g}_\text{S} \partial_x \hat{g}_\text{S} &= [\hat{g}_\text{S},\hat{g}_\text{N}] - \i \frac{G_\phi^\text{S}}{G_T}[\hat{\tau}_3, \hat{g}_\text{S}].
\end{align}
The magnitude $G_\phi^\text{S}$ of the phase-shifts induced in the superconducting region are equal to $G_\phi^\text{N}$ in the absence of a Fermi-vector mismatch, but will in general be different. The normal-state conductivities are given by
\begin{align}
\sigma_\text{N(S)} = \frac{d_\text{N(S)}}{R_\text{N(S)}A},
\end{align}
with $A$ as the interface area and $R_\text{N(S)}$ is the normal-state resistance. Since it is reasonable to assume that the barrier 
region features a higher electrical resistance than the bulk of the materials, we shall set $R_\text{B}/R_\text{S}=4$ in what follows. 
Moreover, we fix the width of the normal layer to $d_\text{N}/\xi_\text{S} = 1.0$.

Due to an inverse proximity effect, the superconductor should also be influenced by the presence of $G_\phi\neq0$, and one expects 
that an odd-frequency triplet component would be induced near the interface on the superconducting side. Therefore, we will also study 
how this inverse proximity effect is manifested in the superconducting DOS. We will focus on the influence of the spin-DIPS $G_\phi$, 
considering an equal magnitude of spin-DIPS in both regions, i.e. $G_\phi^\text{N} = G_\phi^\text{S}$. Consider first a situation 
where the superconducting region acts as a reservoir and is very weakly affected by the proximity effect. To this end, we set 
$d_\text{S}/\xi_\text{S}=5.0$ and  $\sigma_\text{N}/\sigma_\text{S} = 0.2$, ensuring in this way that both $d_\text{S}\gg d_\text{N}$ 
and that the superconducting region is less disordered than the normal region. 

\begin{figure*}
\centering
\resizebox{0.99\textwidth}{!}{
\includegraphics{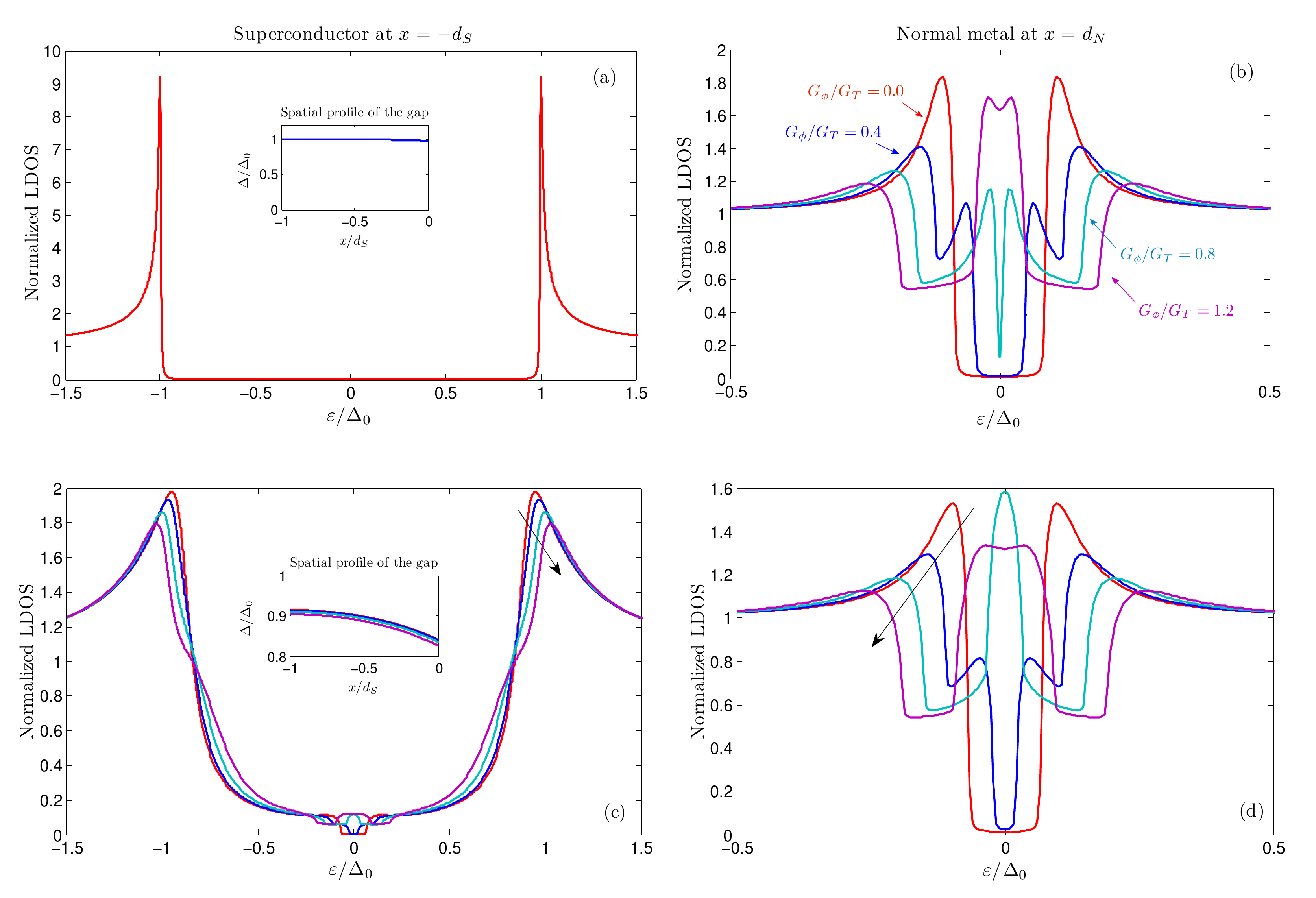}}
\caption{(Color online) (a) and (c): Plot of the DOS in the superconductor at $x=-d_\text{S}$. (b) and (d): the DOS in the normal metal at $x=d_\text{N}$. \textit{Insets:} the spatial profile of the superconducting order parameter. The black arrows indicate an increasing value of $G_\phi/G_T$.
In the top row, we model a scenario where the superconductor acts as a reservoir, we have set $d_\text{S}/\xi_\text{S}=5.0$, 
$d_\text{N}/\xi_\text{S} = 1.0$, $\sigma_\text{N}/\sigma_\text{S} = 0.2$, and $G_\phi^\text{S}=G_\phi^\text{N}\equiv G_\phi$. 
In the bottom row, we model a scenario where the proximity effect is expected to be substantial in both the N and S regions, 
we have set $d_\text{S}/\xi_\text{S}=1.0$,  $d_\text{N}/\xi_\text{S} = 1.0$, $\sigma_\text{N}/\sigma_\text{S} = 1.0$, 
and $G_\phi^\text{S}=G_\phi^\text{N}\equiv G_\phi$. }
\label{fig:DOS_dfive} 
\label{fig:DOS_done} 
\end{figure*}
\par
The results are shown in the top row of Fig. \ref{fig:DOS_dfive}, where we plot the DOS in the superconducting region, the normal metal 
region, and also the spatial depletion of the order parameter. The DOS is plotted at $x=-d_\mathrm{S}$ in the superconducting region 
and $x=d_\mathrm{N}$ in the normal metal region, and may be probed by tunneling spectroscopy measurements through an insulator. In 
the superconducting region, the results are virtually independent of $G_\phi$ in the present case of a reservoir modeled by
$d_\text{S}/\xi_\text{S}=5.0$, so we consider only $G_\phi=0$ there. As seen, both the inverse proximity effect and the gap 
depletion are negligible. However, the DOS in the normal metal region is highly sensitive to the presence of $G_\phi$. In 
particular, the low-energy DOS displays a strong dependence on the ratio $G_\phi/G_T$. We will comment further
on this below.

\par
In the bottom row of Fig. \ref{fig:DOS_done}, we investigate a scenario where the superconducting region no longer acts as a reservoir, 
and where the proximity effect is expected to be substantial in both regions. To this end, we fix  $d_\text{S}/\xi_\text{S}= 1.0$ and $\sigma_\text{N}/\sigma_\text{S} = 1.0$. In this case, the proximity effect in the superconducting region is much stronger than in the 
reservoir case of $d_\text{S}/\xi_\text{S}=5.0$, and the depletion of the  superconducting order parameter is more pronounced. In 
particular, the DOS at Fermi level is no longer zero and depends on the value of $G_\phi$. However, both the DOS and the superconducting 
order parameter remain quite insensitive to a variation in $G_\phi$. In the normal metal region, the behavior is similar to the 
reservoir case, although the peak structure at zero-energy now appears for a lower value of $G_\phi$.

\begin{figure}
\centering
\resizebox{0.40\textwidth}{!}{
\includegraphics{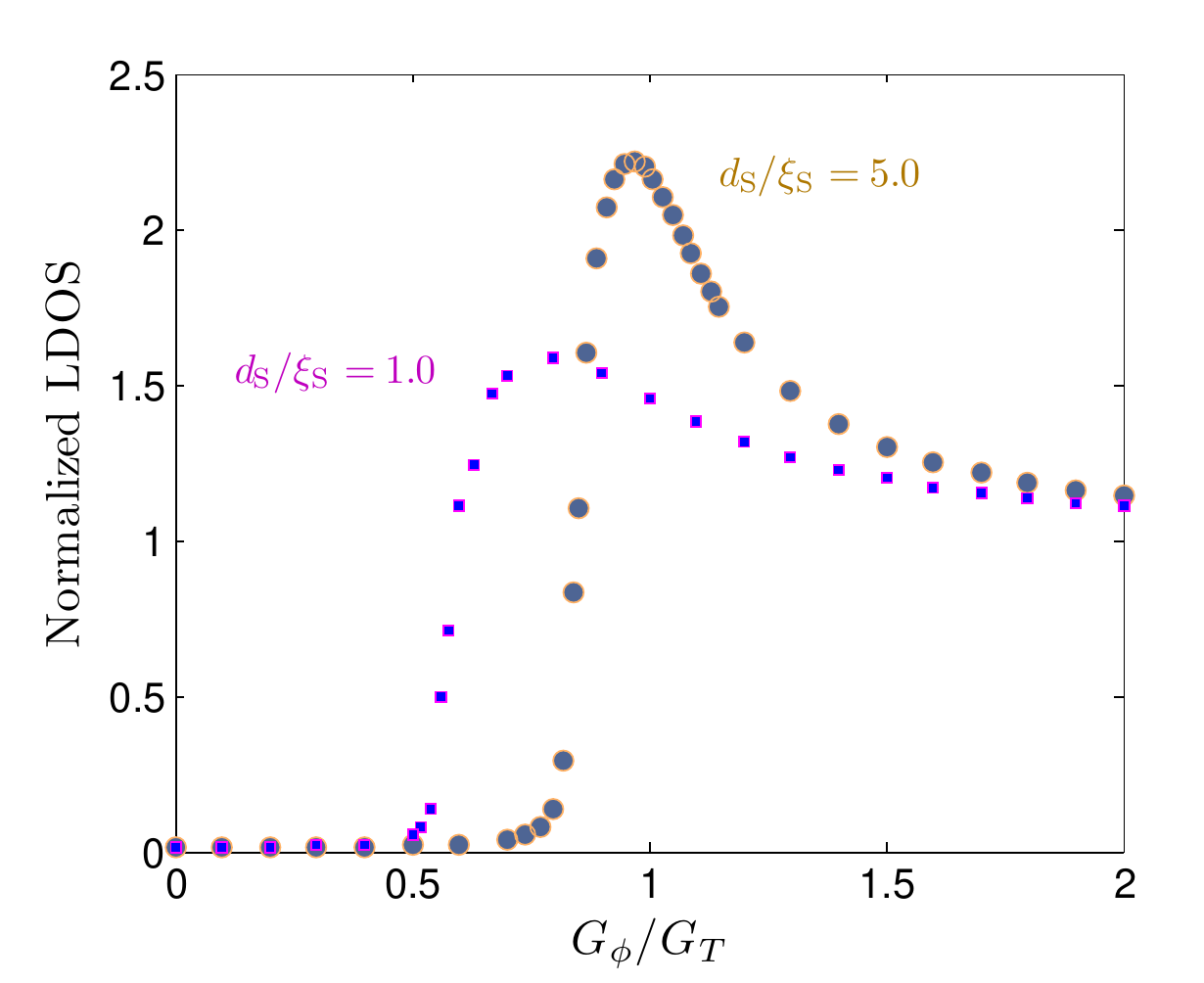}}
\caption{(Color online) Plot of the zero-energy DOS in the normal metal versus $G_\phi/G_T$ for the two cases of $d_\text{S}/\xi_\text{S}=5.0$ (with $\sigma_\text{N}/\sigma_\text{S} = 0.2$) and $d_\text{S}/\xi_\text{S}=1.0$ (with $\sigma_\text{N}/\sigma_\text{S} = 1.0$). As seen, an abrupt transition occurs at a value $G_\phi=\eta G_T$, where $\eta\leq 1$. }
\label{fig:zeroDOS} 
\end{figure}
\begin{figure}
\centering
\resizebox{0.48\textwidth}{!}{
\includegraphics{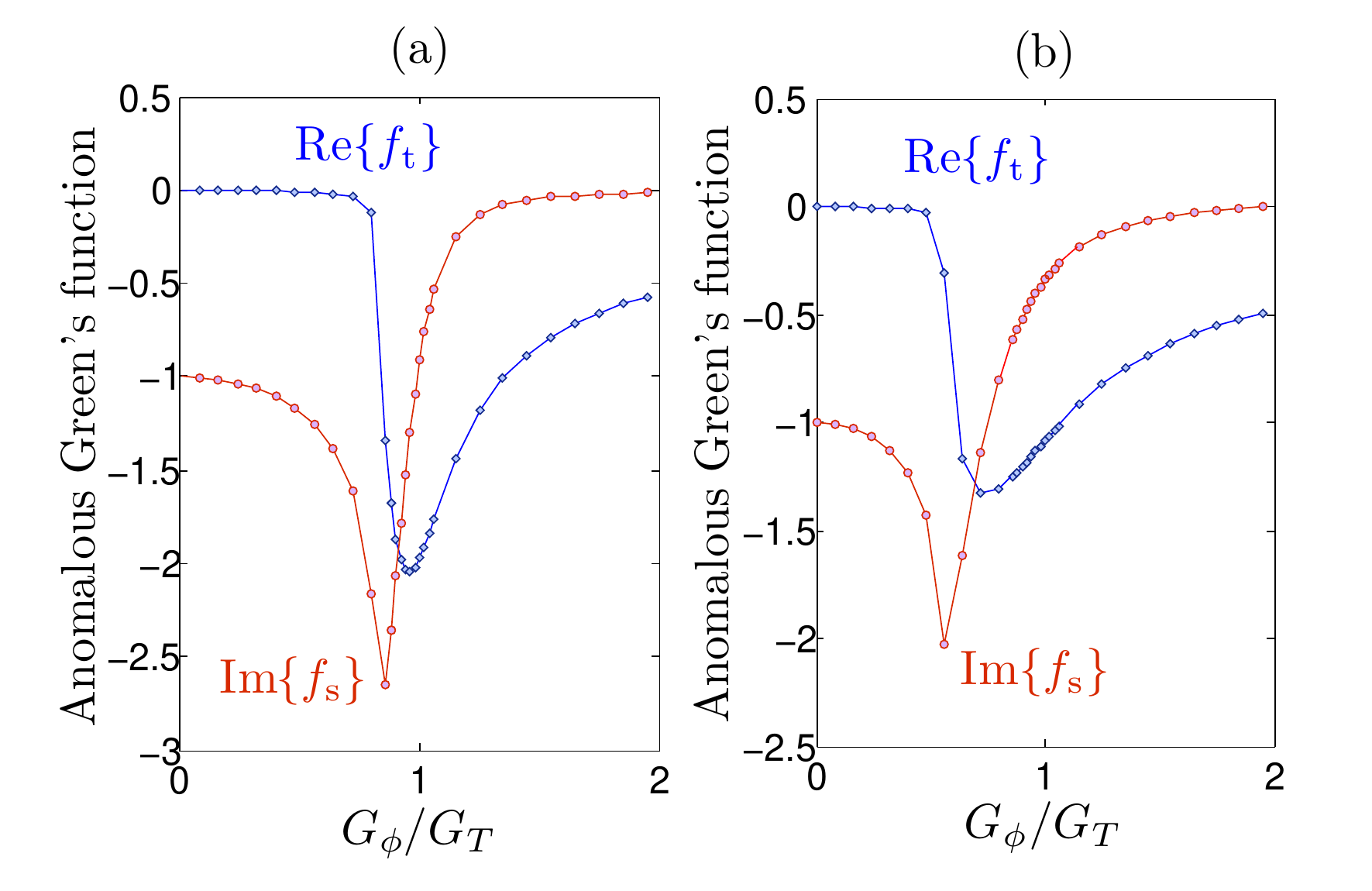}}
\caption{(Color online) Plot of the anomalous Green's function at zero energy (Fermi level) as a function of $G_\phi/G_T$ at $x=d_\text{N}$. 
In (a) we use $d_\text{S}/\xi_\text{S}=5.0, \sigma_\text{N}/\sigma_\text{S} = 0.2$, while in (b) we use $d_\text{S}/\xi_\text{S}=1.0, \sigma_\text{N}/\sigma_\text{S} = 1.0$. As seen, in both cases a transition occurs from singlet to triplet correlations at 
$G_\phi/G_T=\eta$ where $\eta\in\{0,1\}$. We have included inelastic scattering and solved self-consistently for the order 
parameter.}
\label{fig:anomalous} 
\end{figure}
\par
We are particularly interested in seeing if the even- to odd-frequency conversion predicted from the analytical treatment in 
Sec. \ref{sec:diffusive} is equally pronounced in this numerical, self-consistent treatment. To this end, we plot in 
Fig. \ref{fig:zeroDOS} the zero-energy DOS in the normal metal at $x=d_\mathrm{N}$ as a function of $G_\phi/G_T$ for both 
the case of a superconducting reservoir $(d_\text{S}/\xi_\text{S}=5.0, \sigma_\text{N}/\sigma_\text{S} = 0.2)$ and a thin 
layer $(d_\text{S}/\xi_\text{S}=1.0, \sigma_\text{N}/\sigma_\text{S} = 1.0)$. In both cases, the transition from a fully 
suppressed low-energy DOS to an enhanced low-energy DOS appears at 
\begin{align}
G_\phi=\eta G_T,
\end{align}
where $\eta\leq 1$. \textit{This is a clear signature of the transition from pure even- to pure odd-frequency correlations.} The 
corresponding behavior of the anomalous Green's function is shown in Fig. \ref{fig:anomalous}, where we have included inelastic 
scattering and solved self-consistently for the superconducting orer parameter. As seen, the correlations undergo a rapid transition 
from singlet to triplet at $G_\phi/G_T=\eta$, with $\eta\in\{0,1\}$. 

From our above findings, it then follows that the even- to odd-frequency conversion persists also for thin superconducting layers 
comparable in size to the coherence length $\xi_\text{S}$, as long as the inverse proximity effect is relatively weak, with a 
concomitant shift in the critical interface resistance where the pairing transition occurs. 
 
\section{Ballistic limit}\label{sec:ballistic}
 
\subsection{Theory}\label{sec:theory_ball}
Turning our attention now to the ballistic limit, our strategy will be to solve the Eilenberger equation and supplement the solution 
with boundary conditions obtained by means of the $\Smat$-matrix method elaborated upon in a number of
works.\cite{Millis,fogelstrom00,zhao04,eschrig07,eschrig08,eschrig09} The retarded Green's function $\hat{g} \equiv \hat{g}^\text{R}$ 
is in this case most conveniently parameterized by Riccati-amplitudes\cite{nagato,schopohl,eschrig00,cuevas06} $\{\g,\gt\}$, 
where
\begin{align}\label{eq:para_ballistic}
\hat{g} =-\i\pi \begin{pmatrix}
\N\; (\; \underline{1}+\g\; \gt\; ) & 2\; \N\; \g \\
-2\; \Nt\; \gt & -\Nt\; (\; \underline{1} + \gt\; \g\; )\\
\end{pmatrix},
\end{align}
and the normalization matrices read:
\begin{align}
\N = (\; \underline{1}-\g\; \gt\; )^{-1},\; \Nt = (\; \underline{1}-\gt\; \g\; )^{-1}.
\end{align}
Here, we use the notation of Ref.~\onlinecite{eschrig00},  assuming the Green's function to be normalized as $\hat{g}^2 = -\pi^2\hat{1}$. 
The Eilenberger equation for the propagator in the normal region, $\hat g=\hat g_\text{N}$, reads 
\begin{align}
\i v_{Fx} \partial_x \hat g_\text{N} + [\eps \hat \rho_3 ,\hat g_\text{N}] =\hat 0,
\end{align}
where $\hat \rho_3=\diag(1,1,-1,-1)$.
For the boundary conditions at the interface we closely follow 
the $\Smat$-matrix approach in the form presented in Ref.~\onlinecite{eschrig09}.
The scattering approach describes the system by separating it into a scattering region, which cannot be described within quasiclassical 
(QC) theory, and asymptotic regions on both sides of the interface, where QC theory is applicable.\cite{zaitsev84} The scattering region 
must be small compared to the coherence length. It must also extend far enough into the asymptotic region, such that the QC theory is 
applicable. The $\Smat$-matrix approach essentially consists of determining the unknown Riccati amplitudes corresponding to trajectories 
starting at the interface and moving into the bulk on each side by relating them to the known Riccati amplitudes describing trajectories 
starting in the bulk and moving towards the interface. These two sets of amplitudes are related precisely via the $\Smat$-matrix. 
\par
The details of the $\Smat$-matrix depend on what kind of interface is considered. For our purposes, we shall consider a quite 
general model. Namely, an interface which is \textit{(i)} partially transmitting (non-ideal), \textit{(ii)} specular (parallel momentum is 
conserved), and \textit{(iii)} spin-active (giving rise to spin-mixing and spin-filtering effects). The $\Smat$-matrix is evaluated 
at the Fermi level in the quasiclassical approximation, i.e. $\Smat \equiv \Smat(\boldsymbol{p}_\text{F})$, and can be written as
\begin{align}
\Smat &= \begin{pmatrix}
\hat{S}_\text{SS} & \hat{S}_\text{SN} \\
\hat{S}_\text{NS} & \hat{S}_\text{NN} \\
\end{pmatrix}
\equiv \begin{pmatrix}
\hat{R}_\text{S} & \hat{T}_\text{SN} \\
\hat{T}_\text{NS} & -\hat{R}_\text{N} \\
\end{pmatrix}.
\label{Rparameters}
\end{align}
The indices $\text{S}$ and $\text{N}$ refer to the superconducting and normal metallic side of the interface, respectively. Thus, 
$\hat{S}_\text{SS}$ describes reflection processes at the superconducting side of the interface, whereas $\hat{S}_\text{SN}$ 
describes transmission from the superconductor to the normal metal. The elements $\hat{S}_{ij}$ with $\{i,j\} \in \{\text{S},\text{N}\}$ 
are diagonal in particle-hole space according to
\begin{align}
\hat{S}_{ij} &= \begin{pmatrix}
\underline{S}_{ij}(\ppara) & \underline{0}\notag\\
\underline{0} & \underline{S}_{ji}^\text{tr}(-\ppara) \notag\\
\end{pmatrix},
\end{align}
where $\ppara$ denotes the component of the momentum parallel to the interface, and the superscript tr denotes matrix 
transpose. In the presence of an inversion symmetry within the interface plane, the sign of $\ppara$ is unimportant. 
In general, interface scattering may allow for spin-flip processes, namely when 
spin-rotation invariance is completely broken in the system under consideration. 
The details will depend on the micromagnetic properties of the interface.
Here, we will treat the common case that spin-rotation invariance is only partially 
broken, i.e. it is still present with respect to rotations around the axis along 
the magnetic moment of the interface.
Choosing our quantization axis along this direction,
the scattering matrix is also diagonal in spin-space and has the general form: 
\begin{equation} 
\underline{S}_{ij}=\left(\begin{array}{cc} s_{ij\uparrow} e^{i \vartheta_{ij\uparrow}} & 0 \\ 0 & s_{ij\downarrow} e^{i \vartheta_{ij\downarrow}}\end{array}\right).
\end{equation}
Current conservation requires unitarity of the scattering matrix, i.e. the parameters defined by this 
equation are not independent. Moreover, the physical results obtained from quasiclassical theory 
must be gauge invariant in the following sense. 
We may transform the $\mathcal{S}$-matrix by:
\begin{equation}\label{gauge}\hat{\mathcal{S}}'=\left(\begin{array}{cc} e^{i\eta_1/2}\underline{1} & 0 \\ 0 &  e^{i\eta_2/2}\underline{1}\end{array}\right)\hat{\mathcal{S}}\left(\begin{array}{cc} e^{i\eta_1/2}\underline{1} & 0 \\ 0 &  e^{i\eta_2/2}\underline{1}\end{array}\right).\end{equation}
without changing the solutions of the quasiclassical boundary conditions.
This additional gauge freedom is related to the fact that only the envelope of the wave function 
enters quasiclassical quantities sufficiently far away from the interface. 
A transformation according to Eq.~\eqref{gauge} only changes the wavefunction 
on either side of the interface
by a scalar phase factor and thus is irrelevant on the quasiclassical level.
The same gauge freedom can be used to show that the precise definition of which part of the system
is to be included in the scattering region (within the abovementioned restrictions)
does not influence any physical quantity calculated within QC theory (using a general form
of the $\mathcal{S}$-matrix, for the current problem this is shown in Ref.~\onlinecite{eschrig09}).
Exploiting unitarity and the above gauge freedom, we arrive at the following parameterization of the $\mathcal{S}$-matrix
\begin{align}\label{eq:smatrix} \hat{\mathcal{S}}=\left(\begin{array}{cc} \underline{r}\; e^{i\vartheta_\text{S}\; \underline{\sigma}_z/2} & \underline{t}\; e^{i(\vartheta_\text{SN}\; \underline{\sigma}_z+\phi'\; \underline{1})/2} \\
\underline{t}\; e^{i(\vartheta_\text{NS}\; \underline{\sigma}_z-\phi'\; \underline{1})/2} & -\underline{r}\; e^{i\vartheta_\text{N}\; \underline{\sigma}_z/2}             
\end{array}\right)
\end{align}
with 
$\underline{r}=\text{diag}[r_{\uparrow},r_{\downarrow}]$ and $\underline{t}=\text{diag}[t_{\uparrow},t_{\downarrow}]$, and $\underline{\sigma}_z$ is
the third spin Pauli-matrix. Above, $\phi'$ arises due to a possible contribution from a vector potential when a magnetic field is present in the interface region. This contribution is independent of spin, and originates from the time-reversal symmetry breaking by the magnetic field at the interface. It gives an extra phase to the anomalous components, and basically corresponds to the magnetic flux through the interface cross section. It is irrelevant for our purposes, i.e. the behavior of the DOS, but we have kept it for the sake of generality. Unitarity requires
\begin{eqnarray}
t_\sigma^2+r_\sigma^2&=&1, \\
\vartheta_\text{NS}+\vartheta_\text{SN}&=&\vartheta_\text{S}+\vartheta_\text{N}, 
\end{eqnarray}
which implies $6$ free parameters. 

\begin{figure}[t!]
\centering
\resizebox{0.4\textwidth}{!}{
\includegraphics{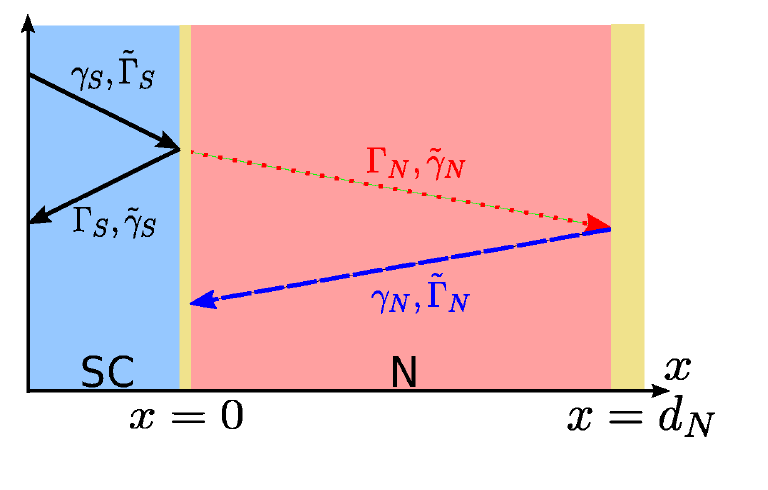}}
\caption{(Color online) Illustration of the incoming and outgoing Riccati-amplitudes in the N$\mid$S bilayer. Lower-case amplitudes $(\gamma,\tilde{\gamma})$ should be integrated towards the interface at $x=0$, whereas upper-case amplitudes $(\Gamma,\tilde{\Gamma})$ 
should be integrated away from the interface at $x=0$. In the superconductor, we use the bulk solution. All amplitudes are 
homogeneous in direction parallel to the interface.}
\label{fig:riccati} 
\end{figure}

With the $\Smat$-matrix in hand, the remaining step is to write down the appropriate boundary conditions which serve as the 
link between the incoming Riccati amplitudes 
\begin{eqnarray}
\underline{\gamma}_\text{N} &\equiv &
\underline{\gamma}_\text{N}(\ve{p}_{||},-p_x,\varepsilon,x), \quad
\underline{\tilde{\gamma}}_\text{N} \equiv 
\underline{\tilde{\gamma}}_\text{N}(\ve{p}_{||},p_x,\varepsilon,x), \quad
\end{eqnarray}
and the outgoing Riccati amplitudes 
\begin{eqnarray}
\underline{\Gamma}_\text{N} &\equiv &
\underline{\Gamma}_\text{N}(\ve{p}_{||},p_x,\varepsilon,x), \quad
\underline{\tilde{\Gamma}}_\text{N} \equiv 
\underline{\tilde{\Gamma}}_\text{N}(\ve{p}_{||},-p_x,\varepsilon,x). \quad
\end{eqnarray}
measured with respect to the S$\mid$N interface (for the notation see Fig. \ref{fig:riccati}). The general solution of the 
Eilenberger equation in the normal metal region reads
\begin{align}
\underline{\Gamma}_\text{N}(x) &= \underline{\Gamma}_\text{N}(0)\e{2\i \varepsilon x/v_{\text{F}x}},
\\
\underline{\tilde{\gamma}}_\text{N}(x) &= \underline{\tilde{\gamma}}_\text{N}(d_\text{N})\e{-2\i\varepsilon (x-d_\text{N})/v_{\text{F}x}}
\end{align}
for the trajectories along $\theta$, whereas for trajectories along $\pi-\theta$ we obtain
\begin{align}
\underline{\gamma}_\text{N}(x) &= \underline{\gamma}_\text{N}(d_\text{N})\e{-2\i\varepsilon(x-d_\text{N})/v_{\text{F}x}},
\\
\underline{\tilde{\Gamma}}_\text{N}(x) &= \underline{\tilde{\Gamma}}_\text{N}(0)\e{2\i\varepsilon x/v_{\text{F}x}},
\end{align}
where we defined $v_{\text{F}x}=v_\text{F}\cos\theta$, and $-\pi/2<\theta<\pi/2$ is assumed.
Here, and in the following, we suppress the parameters $p_x$ and $\varepsilon$ in the argument
list. 
All amplitudes are independent of $\ve{p}_{||}$.
The bulk solution is used for the incoming Riccati amplitudes on the SC-side,
\begin{align}\label{eq:greenSC}
\underline{\gamma}_\text{S} &= -\underline{\tilde{\gamma}}_\text{S} = 
-\; \frac{\Delta_0 }{\varepsilon +i\sqrt{\Delta_0^2-\varepsilon^2}} \; \i \underline{\sigma}_y
\end{align}
where we used a real gauge for the superconducting order parameter $\Delta_0 $,
and $\underline{\sigma}_y$ is the second spin Pauli-matrix.
As shown in Ref.~\onlinecite{eschrig09}, the following boundary conditions at $x=0$ hold:
\begin{align}\label{eq:boundzero}
\underline{\Gamma}_\text{N}(0)&=\underline{\gamma}'_\text{NN}+\underline{\Gamma}_{\text{N}\leftarrow \text{S}}\; \underline{\tilde{\gamma}}_\text{S}(0)\; \underline{\gamma}'_\text{SN}\\\notag
\underline{\Gamma}_{\text{N}\leftarrow \text{S}}&=\underline{\gamma}'_\text{NS}\; \left[\underline{1}-\underline{\tilde{\gamma}}_\text{S}(0)\; \underline{\gamma}_\text{SS}'\right]^{-1}\\\notag
\underline{\gamma}'_{jk}&=\sum_l \underline{S}_{jl}\; \underline{\gamma}_l(0)\; \underline{\tilde{S}}_{lk}
\end{align}
where $j$, $k$ and $l$ run over $\{\text{N},\text{S}\}$.
Analogous equations hold for $\underline{\tilde\Gamma}_\text{N}(0)$.
At $x=d_\text{N}$ we assume perfect and non-spin-active reflection, hence the boundary conditions are trivial:
\begin{equation} 
\underline{\gamma}_\text{N}(d_\text{N})=\underline{\Gamma}_\text{N}(d_\text{N})\equiv \underline{\gamma}_\text{B},\quad\
\underline{\tilde\Gamma}_\text{N}(d_\text{N})=\underline{\tilde \gamma}_\text{N}(d_\text{N})\equiv \underline{\tilde\gamma}_\text{B}
\end{equation}
and result in the following relations between amplitudes at $x=0$ and $x=d_\text{N}$:
\begin{align}\label{eq:boundd}
\underline{\Gamma}_\text{N}(0) &= \underline{\gamma}_\text{B}\e{-2\i\varepsilon d_\text{N}/v_{Fx}},\; 
\underline{\gamma}_\text{N}(0) = \underline{\gamma}_\text{B}\e{2\i\varepsilon d_\text{N}/v_{Fx}}, \\
\underline{\tilde{\Gamma}}_\text{N}(0) &= \underline{\tilde{\gamma}}_\text{B}\e{-2\i\varepsilon d_\text{N}/v_{Fx}},\;
\underline{\tilde{\gamma}}_\text{N}(0) = \underline{\tilde{\gamma}}_\text{B}\e{2\i\varepsilon d_\text{N}/v_{Fx}}.
\end{align}
Replacing $\underline{\Gamma}_\text{N}(0)$ and $\underline{\gamma}_\text{N}(0)$ in Eq.~\eqref{eq:boundzero} according to this relation
yields a quadratic equation in $\gamma_{\text{B},\sigma}$ whose solutions can be determined analytically.\cite{eschrig09}

\subsection{Results}\label{sec:results_ball}

The odd-even frequency conversion which was shown to take place in the diffusive limit also occurs in the ballistic limit, as we show in the following. 
In this case, we
obtain the retarded Green's function using the formalism described in 
Sec. \ref{sec:theory_ball}.
There we derived an equation from Eqs. \eqref{eq:boundzero} and \eqref{eq:boundd} for the Riccati amplitudes in the normal metal region that determine the proximity amplitudes. Following Ref.~\onlinecite{eschrig09}, we obtain analytical expressions for the anomalous 
Green's function in the N region.
The energy-resolved DOS at the outer boundary of the normal layer 
can then directly be obtained from the Riccati amplitudes via:
\begin{equation} \frac{N(\varepsilon)}{N_0}=-\text{Im}\frac{\text{Tr}}{2\pi}
\langle\hat{g}_\text{B}\rangle=\frac{\text{Tr}}{2}\left\langle
(\underline{1}-\underline{\gamma}_\text{B}\underline{\tilde{\gamma}}_\text{B})^{-1}
(\underline{1}+\underline{\gamma}_\text{B}\underline{\tilde{\gamma}}_\text{B})
\right\rangle\end{equation}
where $\langle\bullet\rangle$ denotes the Fermi-surface average given by:
\begin{equation} \langle \bullet\rangle=\frac{1}{N_0}\int_{FS} \frac{d^2 p'_{\mathrm{F}}}{(2\pi\hbar)^3|\ve{v}_{\mathrm{F}}(p'_{\mathrm{F}})|} (\bullet ),
\end{equation}
with the local density of states in the normal state,
\begin{equation} 
N_0=\int_{FS} \frac{d^2 p'_{\mathrm{F}}}{(2\pi\hbar)^3|\ve{v}_{\mathrm{F}}(p'_{\mathrm{F}})|}.
\end{equation}

It is important to realize that only the singlet and the $S_z=0$ triplet component (in a basis where the $z$-axis is along the quantization axis) will be induced in the normal part of the system, since the magnetization of the barrier is uniaxial and has no inhomogeneity. We may then write
\begin{align}
\underline{\gamma}_\text{B} = \begin{pmatrix}
0 & \gamma_+\\
-\gamma_- & 0 \\
\end{pmatrix} =
\begin{pmatrix}
\gamma_+ &0\\
0&\gamma_- \\
\end{pmatrix} \; \i \underline{\sigma}_y
\end{align}
which can be inserted into Eq. (\ref{eq:boundzero}), and similarly for $\underline{\tilde{\gamma}}_\text{B}$. Using the scattering 
matrix defined in Eq.~\eqref{eq:smatrix} and focusing on subgap energies $|\varepsilon|<\Delta_0$, where 
$\gamma_\text{S}=\i\e{\i\Psi}$ with $\Psi=\text{arcsin}(\varepsilon/\Delta_0)$
(here $-\pi/2<\Psi\le \pi/2$),
we get two decoupled equations for $\gamma_\sigma$:
\begin{align}\label{eq:gamma}
\gamma_\sigma^2\e{2\i\phi'} &+ \frac{2u_\sigma }{t_\uparrow t_\downarrow}\gamma_\sigma \e{\i\phi'}  + 1 = 0,
\end{align}
where $\sigma\in \{+,-\}$, and
the function $u_\sigma (\varepsilon )$ is defined as
\begin{align}\label{eq:u}
u_\sigma (\varepsilon) &= \sin\left(\frac{2\varepsilon d_\text{N}}{v_{Fx}} + \sigma\vartheta_+ + \Psi \right) 
\notag\\& \qquad
+r_\uparrow r_\downarrow \sin\left(\frac{2\varepsilon d_\text{N}}{v_{Fx}} + \sigma\vartheta_- - \Psi\right).
\end{align}
Here, we have defined 
$\vartheta_\pm = \frac{1}{2}(\vartheta_{N}\pm \vartheta_{S})$, and the
variable $\sigma $ is to be understood as a factor $\pm 1$ for $\sigma=\pm$.
Eq. (\ref{eq:gamma}) is solved by:
\begin{align}
\gamma_\sigma = \e{-\i\phi'} \left(-\, \frac{u_\sigma}{t_\uparrow t_\downarrow} \pm \sqrt{\frac{u_\sigma^2}{(t_\uparrow t_\downarrow )^2} - 1}\right).
\end{align}
We can write down an equation analogous to Eq. (\ref{eq:gamma}) for 
$\tilde\gamma_\sigma $,
\begin{align}\label{eq:tildegamma}
\tilde\gamma_\sigma^2\e{-2\i\phi'} &+ \frac{2\tilde u_\sigma }{t_\uparrow t_\downarrow}\gamma_\sigma \e{-\i\phi'}  + 1 = 0,
\end{align}
with $\tilde u_\sigma (\varepsilon ) = u_\sigma (-\varepsilon )$.
Noting that $u_-(-\varepsilon)=-u_+ (\varepsilon)$, it follows
that $\tilde \gamma_- = - \gamma_+ \e{2\i \phi'}$.
The correct sign is obtained by requiring (i) that the symmetry relation
holds between $\gamma_\sigma $ and $\tilde \gamma_\sigma $, and (ii) that the 
momentum and spin resolved density of states, e.g.
\begin{equation} 
\frac{N_\uparrow }{N_0}=
\frac{1-\gamma_+\tilde\gamma_- }{1+\gamma_+\tilde\gamma_-}=
\frac{1+\gamma_+^2 \e{2\i\phi' }}{1-\gamma_+^2 \e{2\i\phi' }},
\end{equation}
must be positive.
Also, we must demand $\gamma_\sigma \to 0$ when $t_\uparrow t_\downarrow \to 0$, as in that case the two regions become completely decoupled and the proximity effect should be zero. 
It follows, that the appropriate solution is
\begin{align}
\gamma_\sigma = \e{-\i\phi'}\text{sgn}[u_\sigma] \Big(-\frac{|u_\sigma|}{t_\uparrow t_\downarrow} + 
\sqrt{\frac{u_\sigma^2}{(t_\uparrow t_\downarrow )^2} - 1}\Big).
\end{align}

Having obtained the correct solution for the $\gamma_\sigma$ and $\tilde{\gamma}_\sigma$ quantities, we turn to the anomalous Green's functions. From the parameterization of the Green's function Eq. (\ref{eq:para_ballistic}), we identify
\begin{align}
f_\sigma = -2\pi \i \; \frac{\gamma_\sigma}{1+\gamma_\sigma \tilde{\gamma}_{-\sigma}}= -2\pi \i \; \frac{\gamma_\sigma}{1-\gamma_\sigma^2 \e{2i\phi' }},
\end{align}
where we have defined $f_\pm = f_{\rm s} \pm f_{\rm t}$.
Inserting the solutions of $\gamma_\sigma$ and $\tilde{\gamma}_\sigma$ from above, we obtain 
\begin{align}\label{eq:anomalous1}
f_\sigma(\varepsilon ) = \i\pi \frac{t_\uparrow t_\downarrow \e{-\i\phi'}\text{sgn}[u_\sigma(\varepsilon)]}{\sqrt{u_\sigma(\varepsilon)^2 - (t_\uparrow t_\downarrow)^2}}.
\end{align}

At this point, it is instructive to consider the angular dependence of the Green's function. From the boundary conditions in Sec. \ref{sec:theory_ball} and the general form of $\underline{\gamma}_\text{N}$ and $\underline{\Gamma}_\text{N}$, it is seen that $\underline{\gamma}_\text{N}(d_\text{N}) = \underline{\Gamma}_\text{N}(d_\text{N})$ for all values of $\theta$. In effect, this means that Eq. (\ref{eq:anomalous1}) is valid for any value $\theta$ although we used e.g. $\underline{\gamma}_\text{N}(d_\text{N})$ to obtain them, which only is defined in the range $-\pi/2<\theta<\pi/2$.  It is also worth to note that at the outer surface of the bilayer, all singlet components are even-frequency while all triplet components are odd-frequency. While this can be shown analytically from the above equations, one may also understand it intuitively from the fact that one has specular reflection at the outer surface such that all components must be even in momentum there.

We obtain the energy-resolved DOS at the outer boundary of the normal layer 
from the equations presented above, 
\begin{equation} \frac{N(\varepsilon)}{N_0}=\text{Re}\sum_{\sigma}\left\langle 
\frac{|u_\sigma(\varepsilon )|}{\sqrt{u_\sigma(\varepsilon)^2-(t_\uparrow t_\downarrow)^2}}\right\rangle, \end{equation}
and the pairing amplitudes from Eq.~\eqref{eq:anomalous1}.
In order to investigate the even- to odd-frequency conversion at the chemical potential, 
we are in particular interested in their value at $\varepsilon=0$. 
We obtain for $|u_0|>t_\uparrow t_\downarrow$:
\begin{align}\label{eq:singlet}
f_{\rm s} (\varepsilon=0)= 0,\quad f_{\rm t} (\varepsilon=0)= \i\pi \frac{t_\uparrow t_\downarrow \e{-\i\phi'}\text{sgn}(u_0)}{\sqrt{u_0^2-(t_\uparrow t_\downarrow)^2}},
\end{align}
whereas for $|u_0|<t_\uparrow t_\downarrow$:
\begin{align}\label{eq:triplet}
f_{\rm s} (\varepsilon=0)= \pi \frac{t_\uparrow t_\downarrow \e{-\i\phi'}}{\sqrt{(t_\uparrow t_\downarrow)^2 -u_0^2}},\quad f_{\rm t} (\varepsilon=0)=0.
\end{align}
Here, the parameter $u_0$ is given by
\begin{equation}
\label{u0}
u_0=\sin\left( \frac{\vartheta_\text{N}+\vartheta_\text{S}}{2}\right) +
r_{\uparrow }r_{\downarrow }
\sin\left( \frac{\vartheta_\text{N}-\vartheta_\text{S}}{2}\right).
\end{equation}
In the case that both $\vartheta_\text{N}$ and
$\vartheta_\text{S}$ are of order of
$t_\uparrow t_\downarrow$, and the system is at the same time in the tunneling limit,
we can expand all quantities up to $(t_\uparrow t_\downarrow)^2$ and
thus recover the results of Ref.~\onlinecite{linder_prl_09} that $u_0$ in the above
expressions for the pair amplitudes and for the density of states
can be replaced by $\vartheta_\text{N}$.
Note that the scalar phase $\phi'$ was set to zero in Ref.~\onlinecite{linder_prl_09}, as it has no consequence for the behavior of the DOS. 

\begin{figure}
\includegraphics[width=0.5\columnwidth]{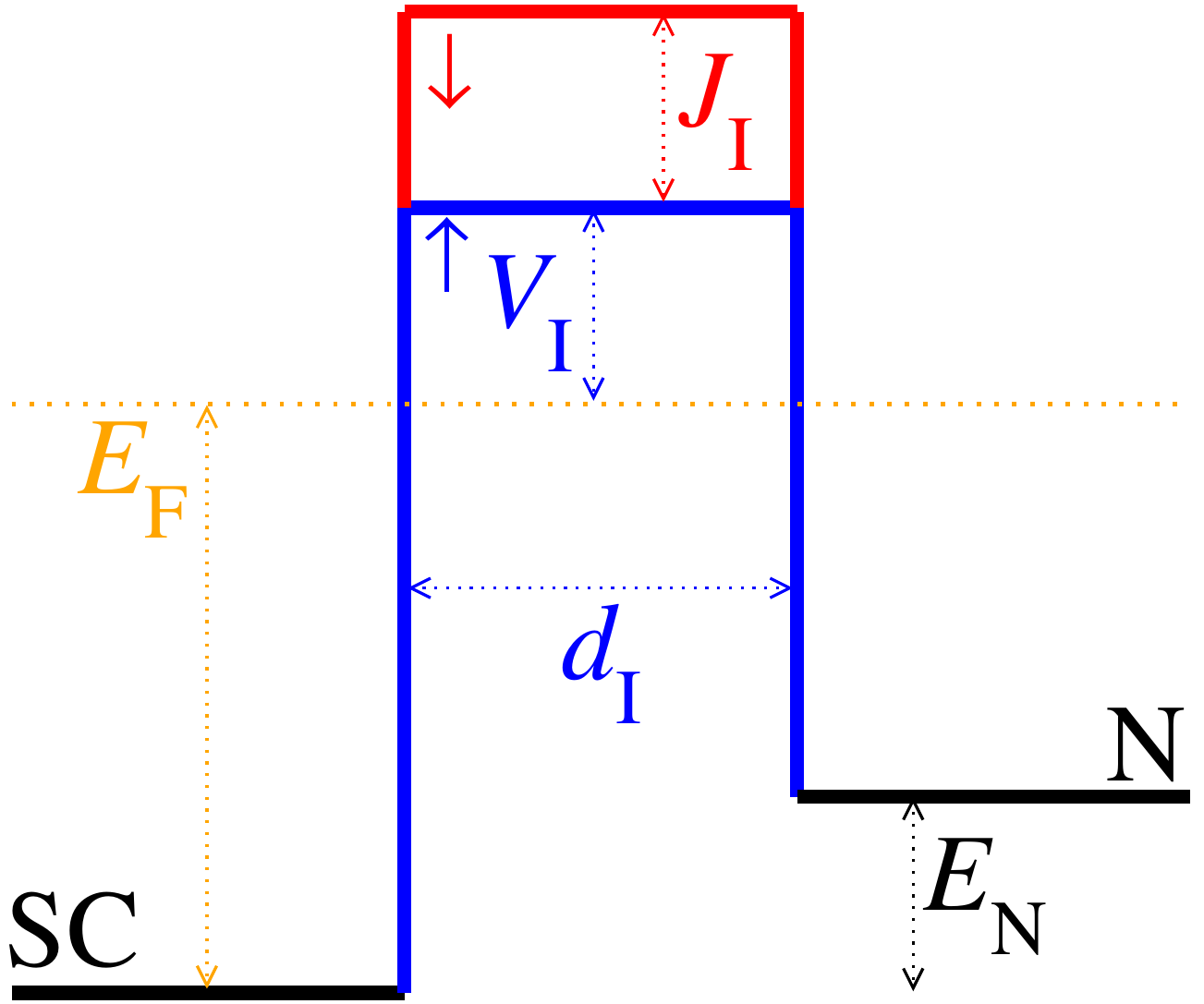}
\caption{
\label{fig:profile}
(color online) Definition of the parameters for the spin-dependent barrier.
For Figs.~\ref{fig:cleanDOS01}-\ref{fig:cleanDOS10} we use 
$V_\text{I}=0.2 E_\text{F}$, and $d_\text{I}=2/k_{\text{F},\text{S}}$.
}
\end{figure}

Considering a realistic interface, it is clear that both the transmission coefficients $t_\sigma$ and the spin-mixing angle $\vartheta_\text{N}$ depend on the angle of incidence $\theta$. 
A systematic study of this angular dependence was performed in Ref~\onlinecite{greinPRB}. While the expressions Eqs.~\eqref{eq:singlet} and \eqref{eq:triplet} are valid even for angle-dependent quantities $t_\sigma$ and $\vartheta_\text{N}$, the Fermi-surface average can in fact add the possibility of a
simultaneous presence of both triplet and singlet correlations at zero energy originating from different incidence angles $\theta $; i.e. 
some trajectories may contribute to the singlet component while others contribute to the triplet. 
In order to discuss the conditions under which that happens,
we performed calculations for an interface layer modeled by a
spin-active box-shaped potential of width $d_\text{I}$, with $d_\text{I}\sim \lambda_\text{F}$. 
Its height is spin-dependent and given by 
$U_{\text{I}\uparrow}= V_\text{I}$,
$U_{\text{I}\downarrow}= V_\text{I}+J_\text{I}$,
where $J_\text{I}$ is the interface exchange field (see Fig.~\ref{fig:profile} for the notation). 

\begin{figure*}
\includegraphics[height=0.65\columnwidth]{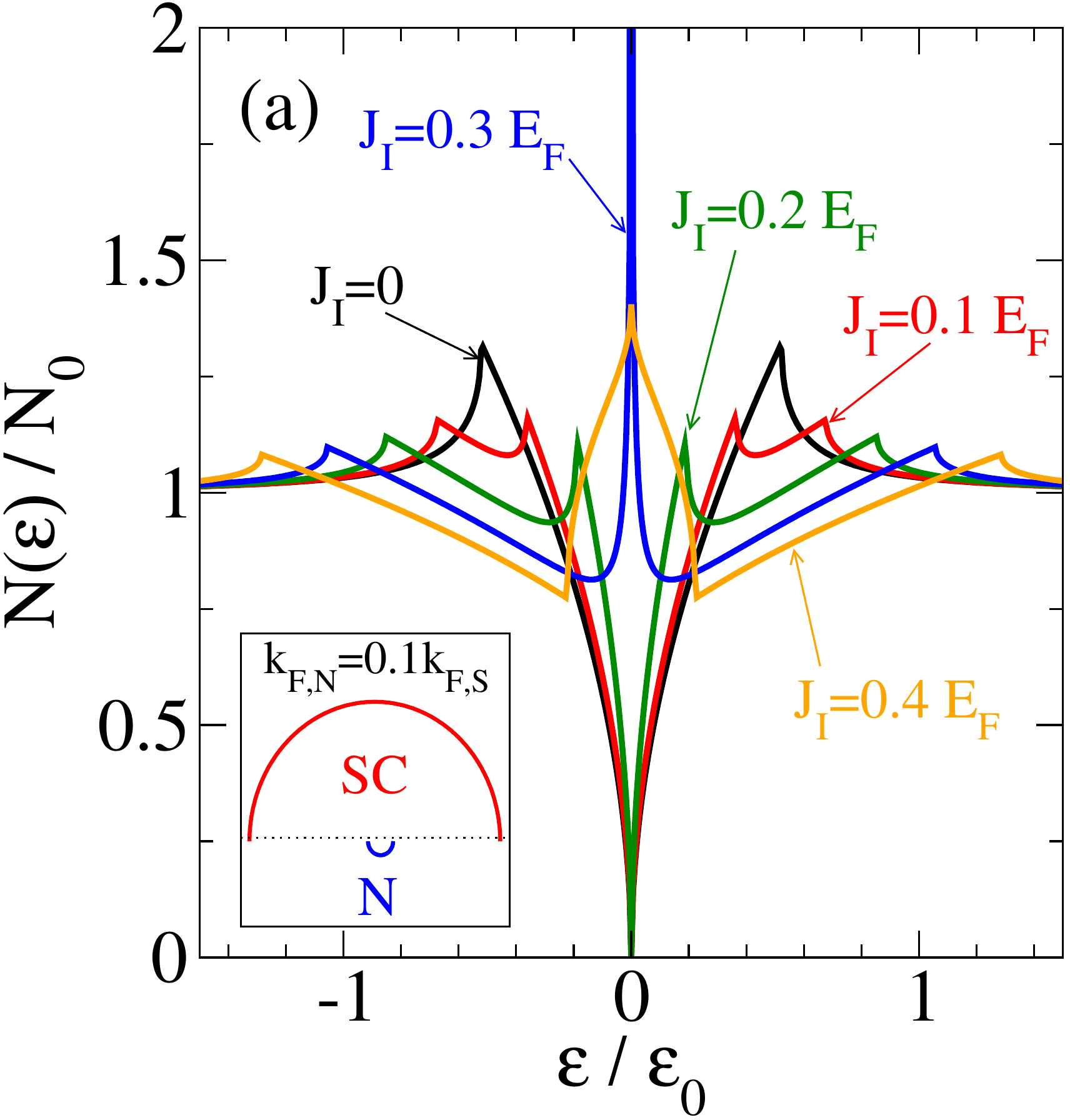}
\includegraphics[height=0.65\columnwidth]{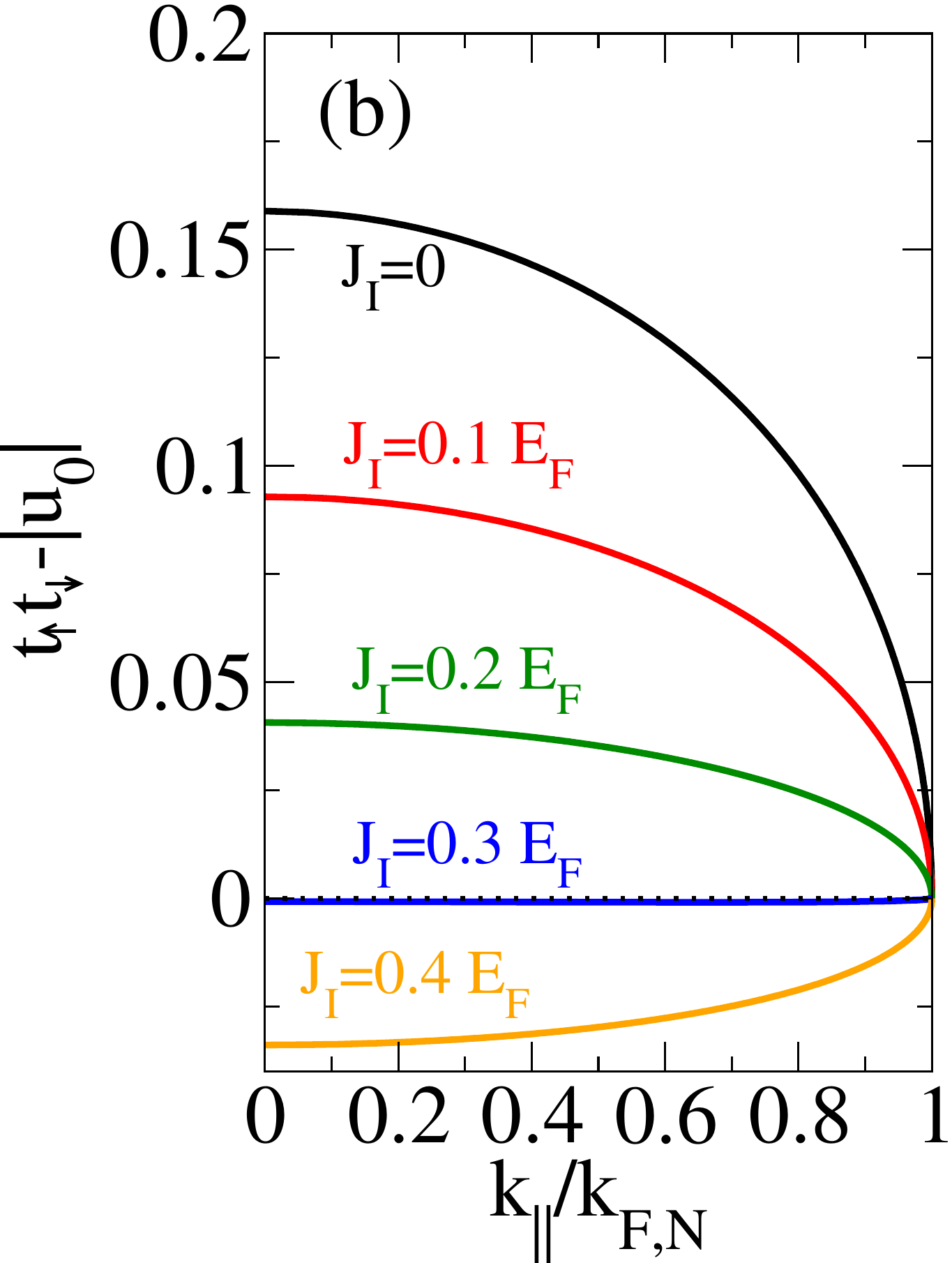}
\includegraphics[height=0.65\columnwidth]{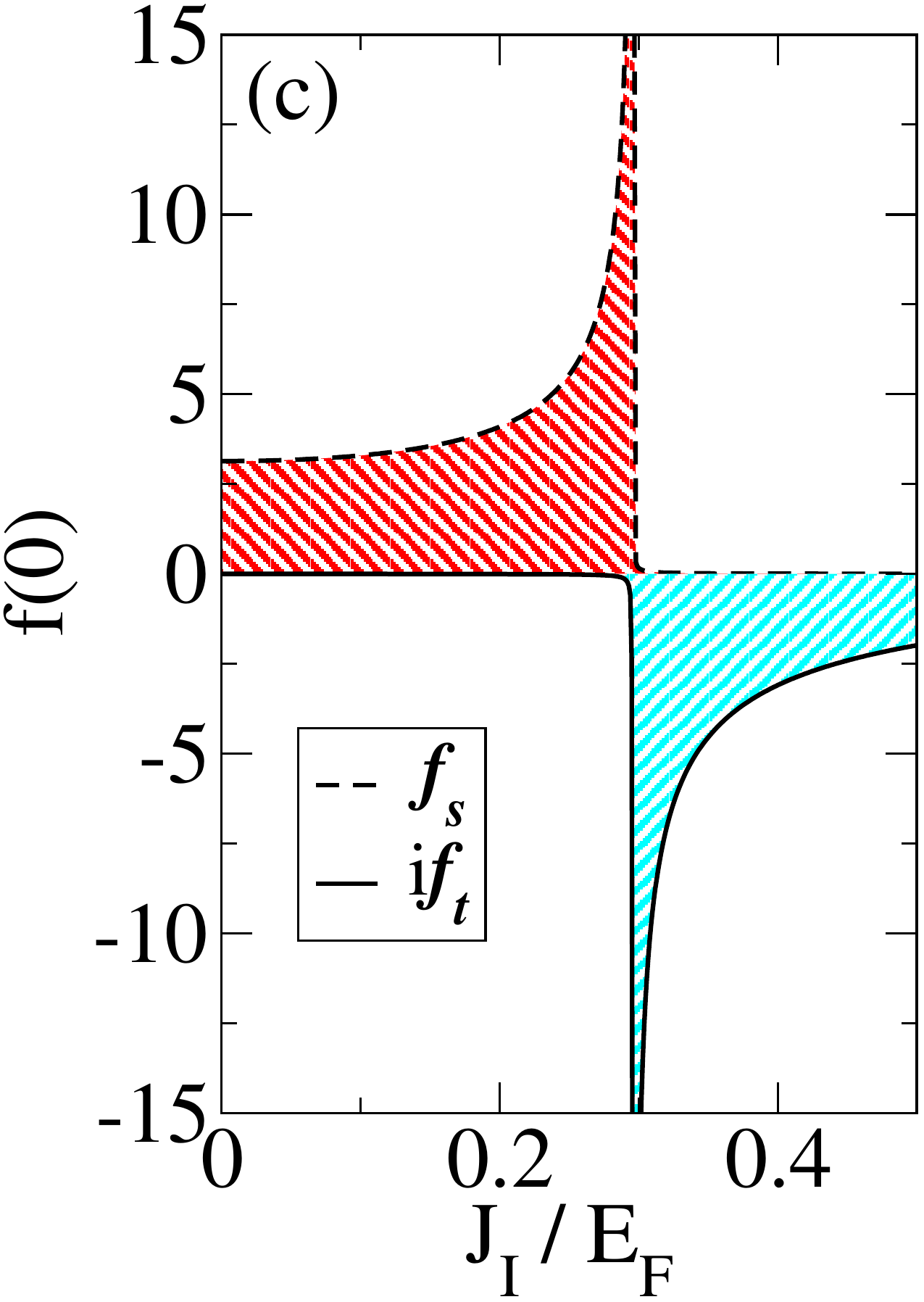}
\includegraphics[height=0.65\columnwidth]{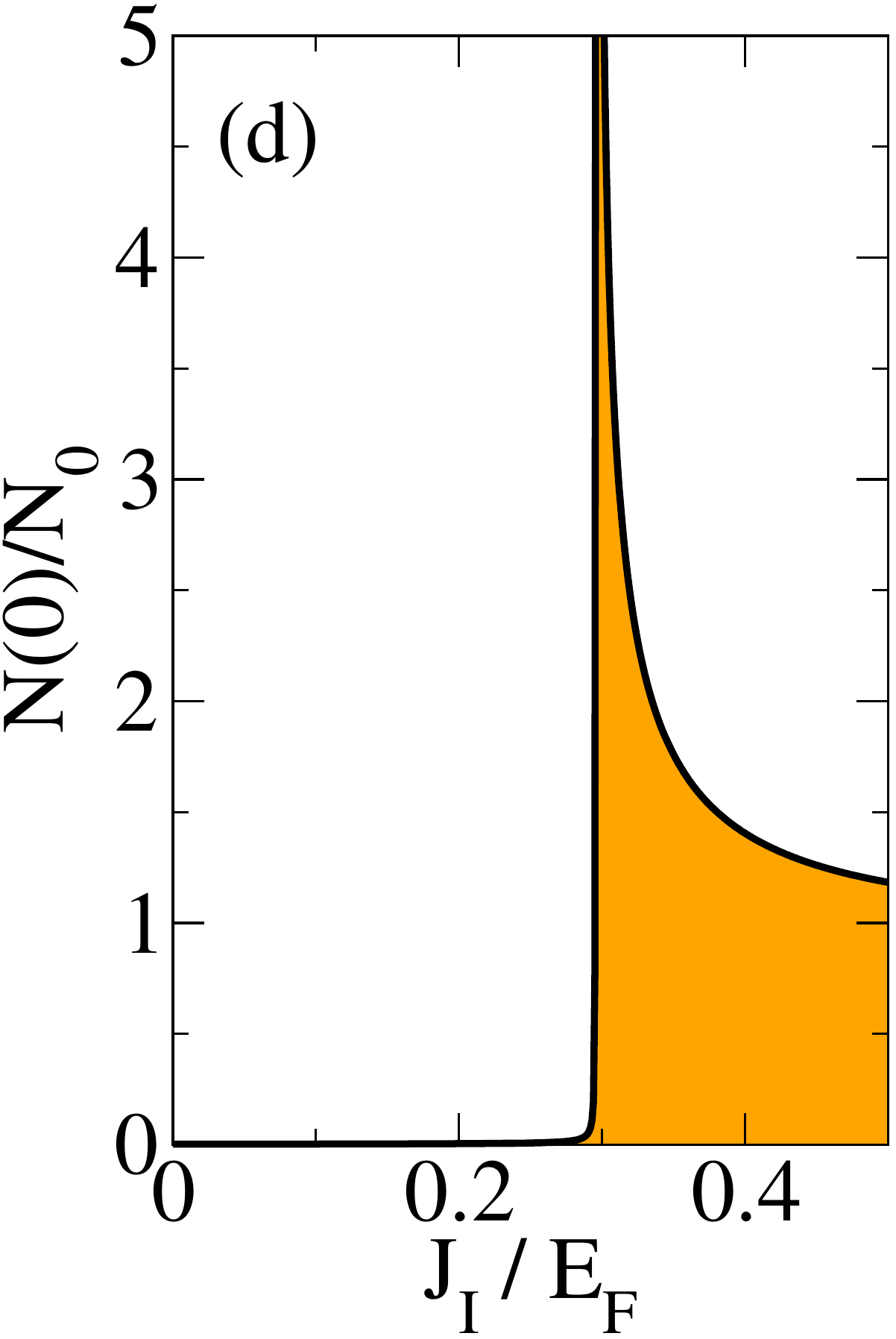}
\caption{(Color online) 
(a) Energy-resolved DOS in the normal metal for different values of the interface exchange field $J_{\rm I}$. 
The energy scale is $\varepsilon_0=[t_{\uparrow}t_{\downarrow} \varepsilon_{\rm Th}](k_{||}=0)$, 
with the Thouless energy $\varepsilon_{\rm Th}=\hbar v_{\rm F,N}/2d_{\rm N}$.
(b)
Interface parameter $t_{\uparrow}t_{\downarrow}-|u_0|$ as a function of trajectory impact (parameterized by $k_{||}$). 
(c)
Singlet and triplet component of the anomalous Green's function at $\varepsilon=0$ as a function of $J_{\rm I}$. 
(d) Density of states at the Fermi level, $N(0)$. 
The Fermi surface mismatch is $k_{\rm F,N}=0.1 k_{\rm F,S}$. The inset in the lower left corner of panel (a)
is meant to illustrate the Fermi-surface mismatch.
In (a)-(d), the interlayer thickness is $d_\text{I}=2 \lambda_{\mathrm{F}}/2\pi$,
and the interface potential $V_{\rm I}=0.2 E_{\rm F}$. 
The width of the normal layer is $d_{\rm N}=\hbar v_{\rm F,N} /\Delta_0$. 
\label{fig:cleanDOS01}}
\end{figure*}
\begin{figure*}
\includegraphics[height=0.65\columnwidth]{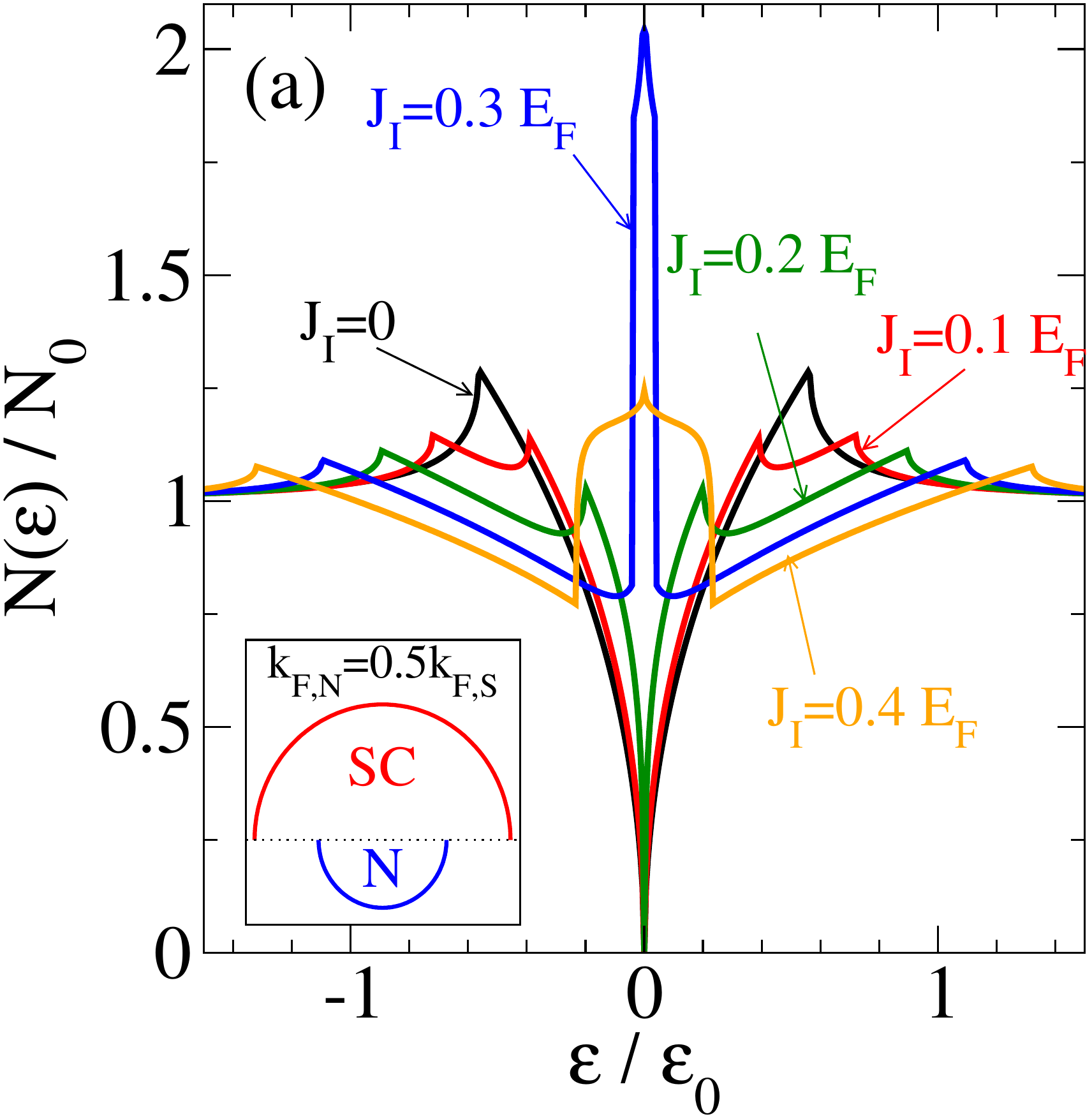}
\includegraphics[height=0.65\columnwidth]{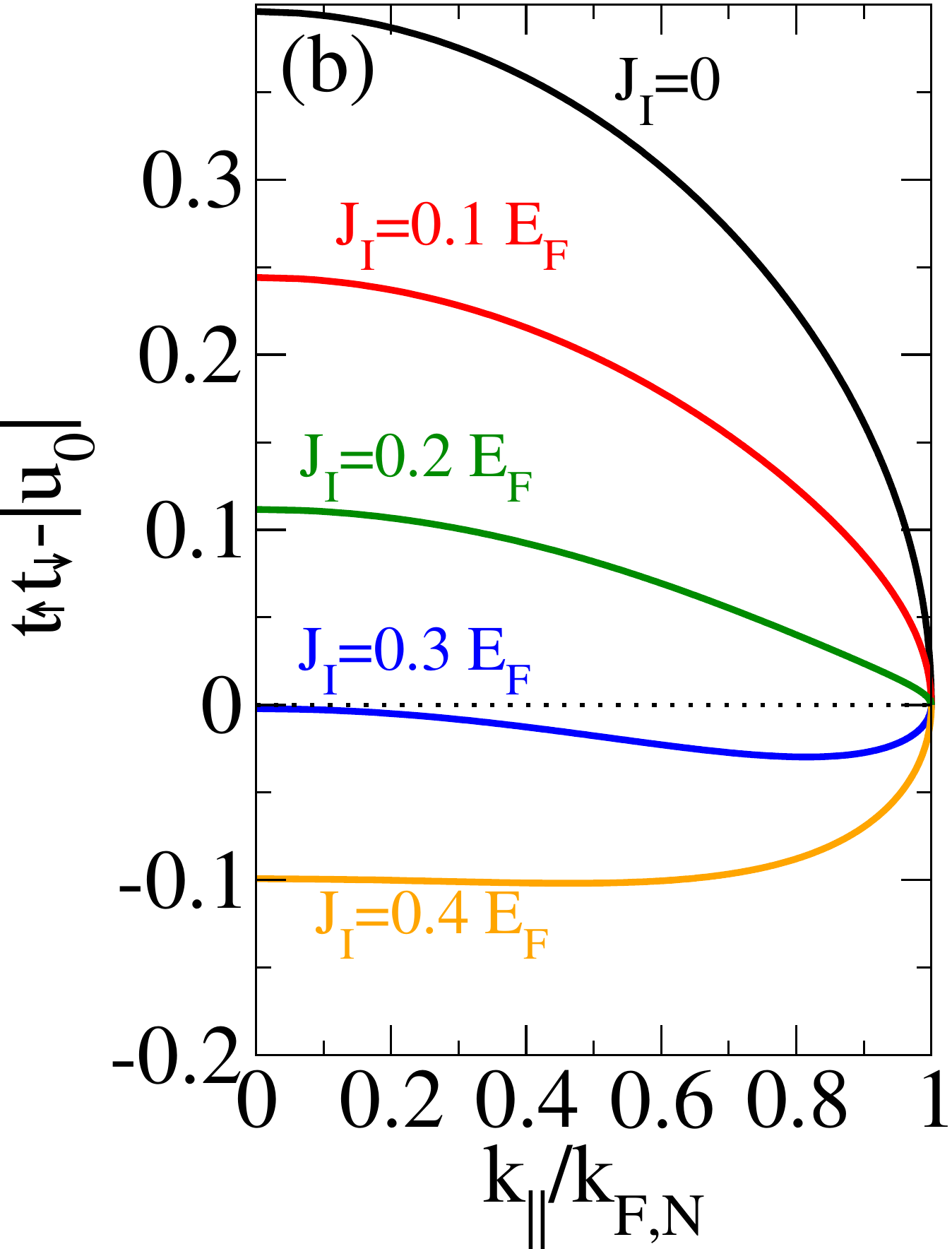}
\includegraphics[height=0.65\columnwidth]{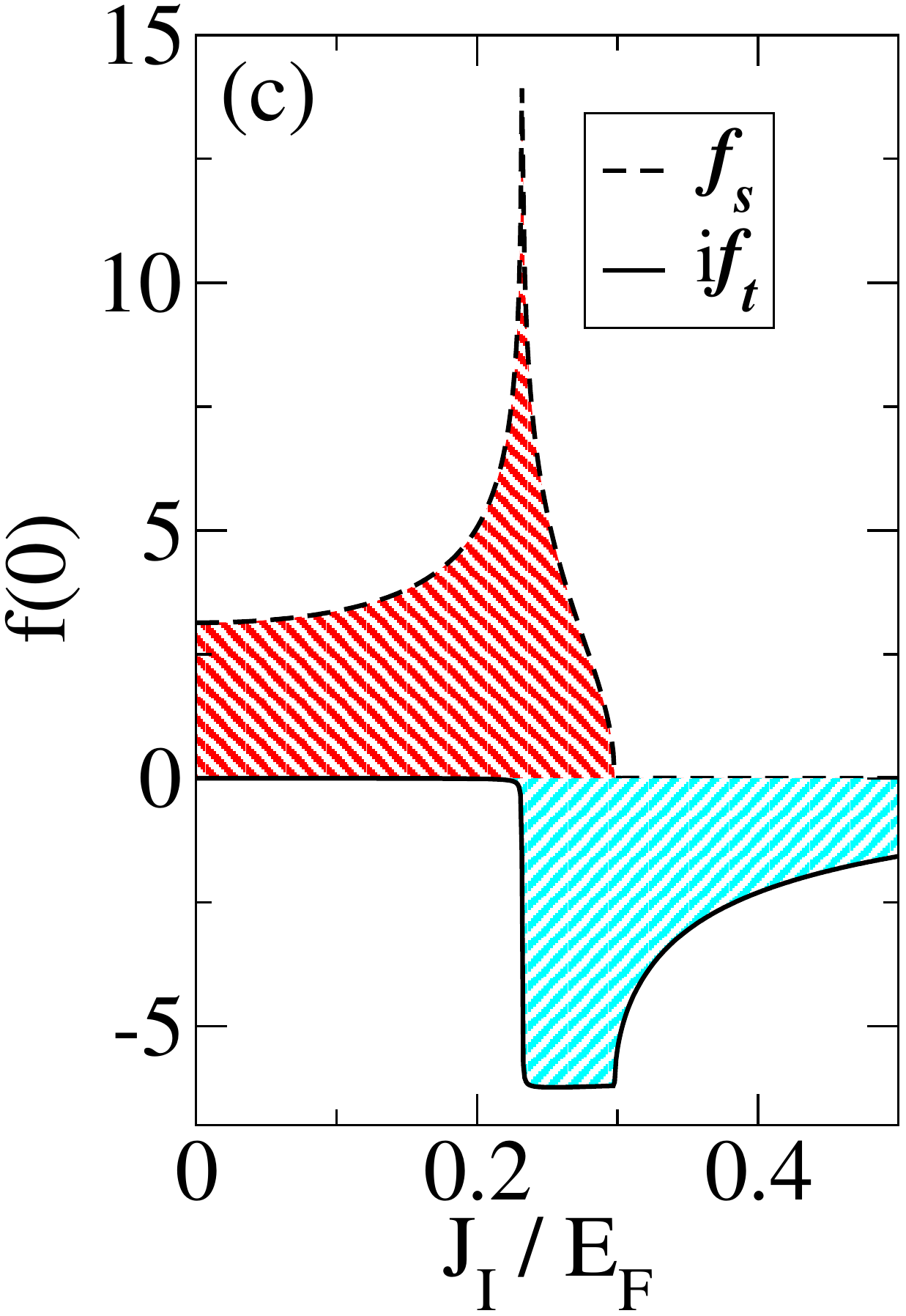}
\includegraphics[height=0.65\columnwidth]{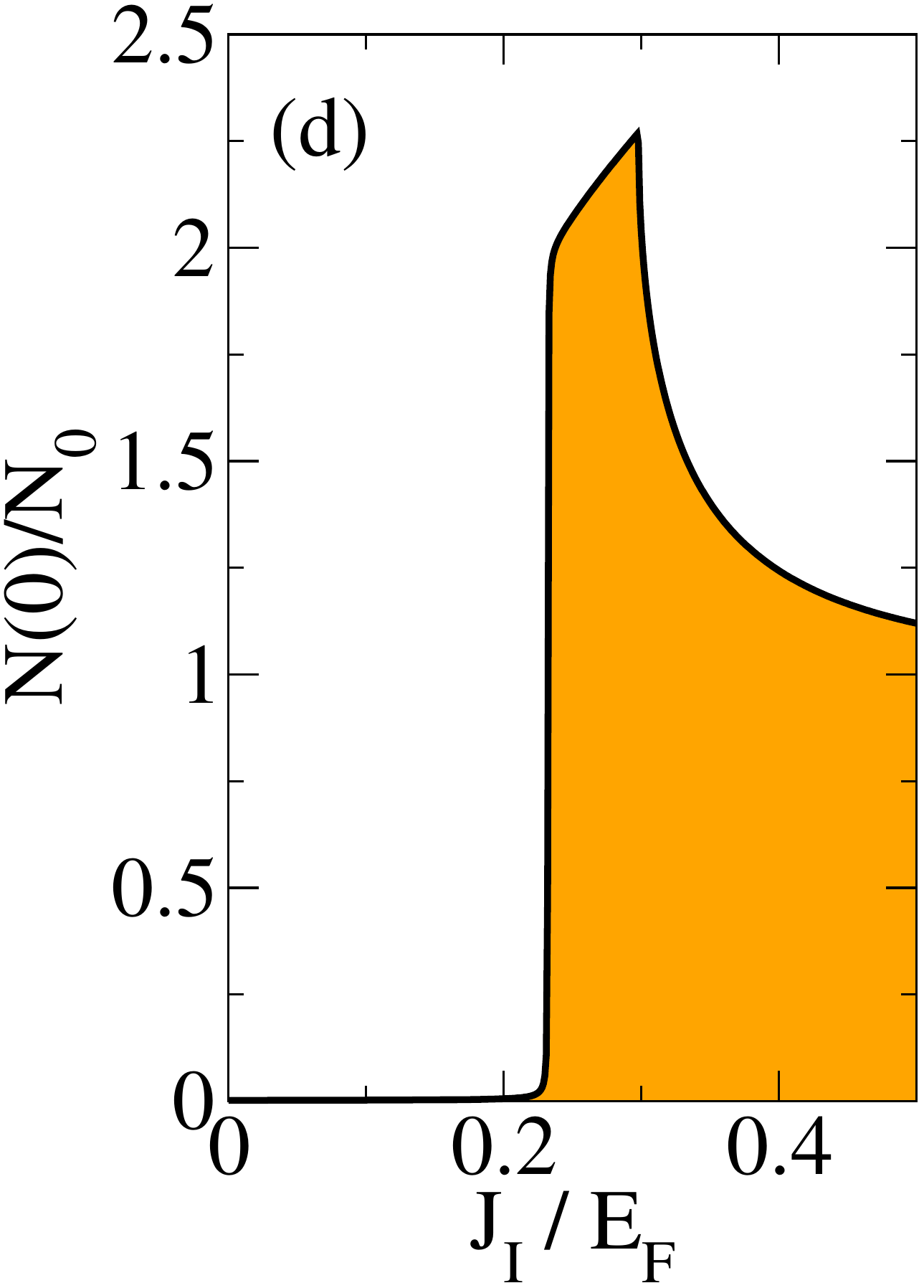}
\caption{(Color online) 
The same as in Fig.~\ref{fig:cleanDOS01}, however
with a different Fermi surface mismatch, $k_{\rm F,N}=0.5 k_{\rm F,S}$.
\label{fig:cleanDOS05}}
\end{figure*}
\begin{figure*}
\includegraphics[height=0.64\columnwidth]{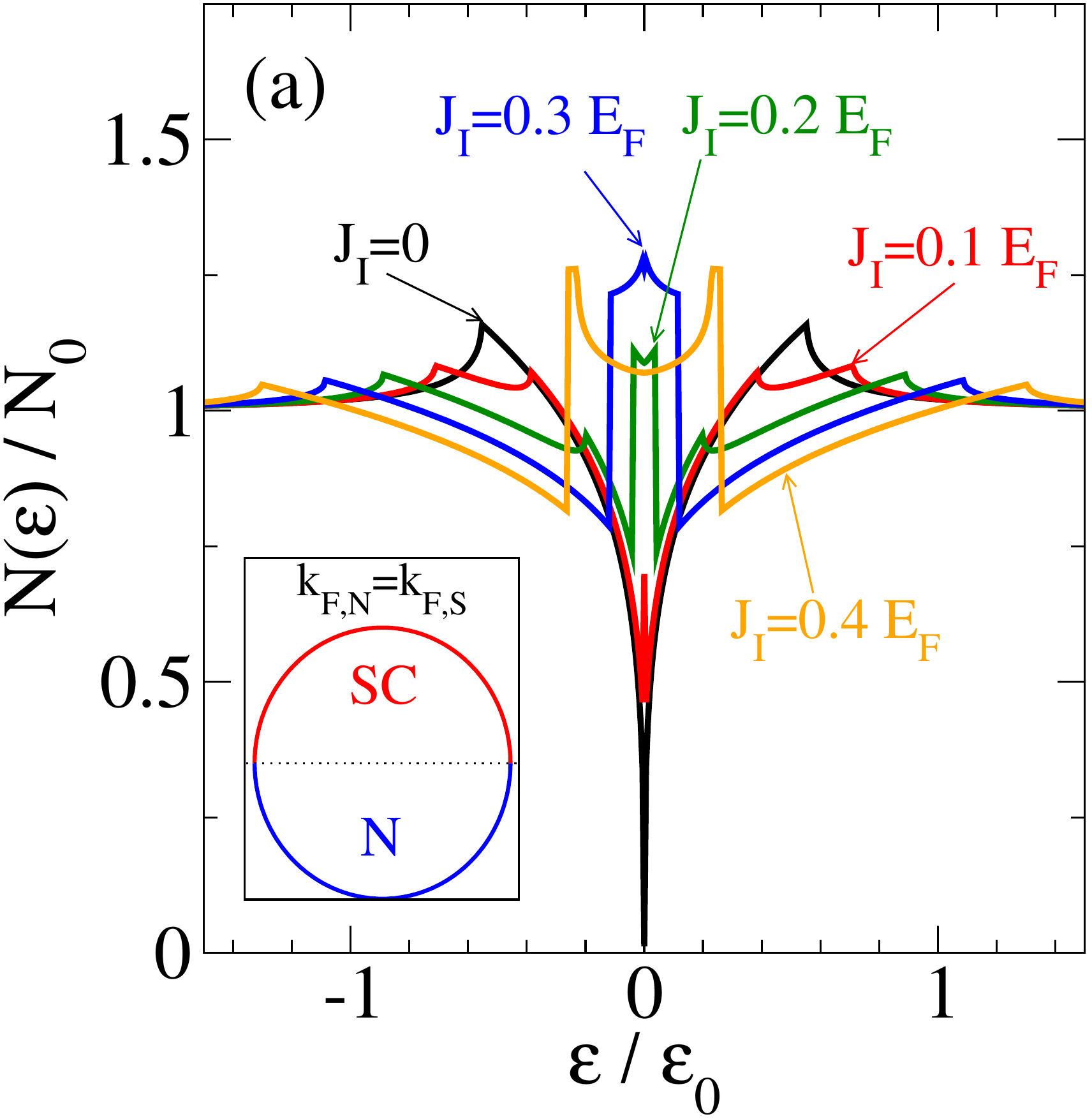}
\includegraphics[height=0.64\columnwidth]{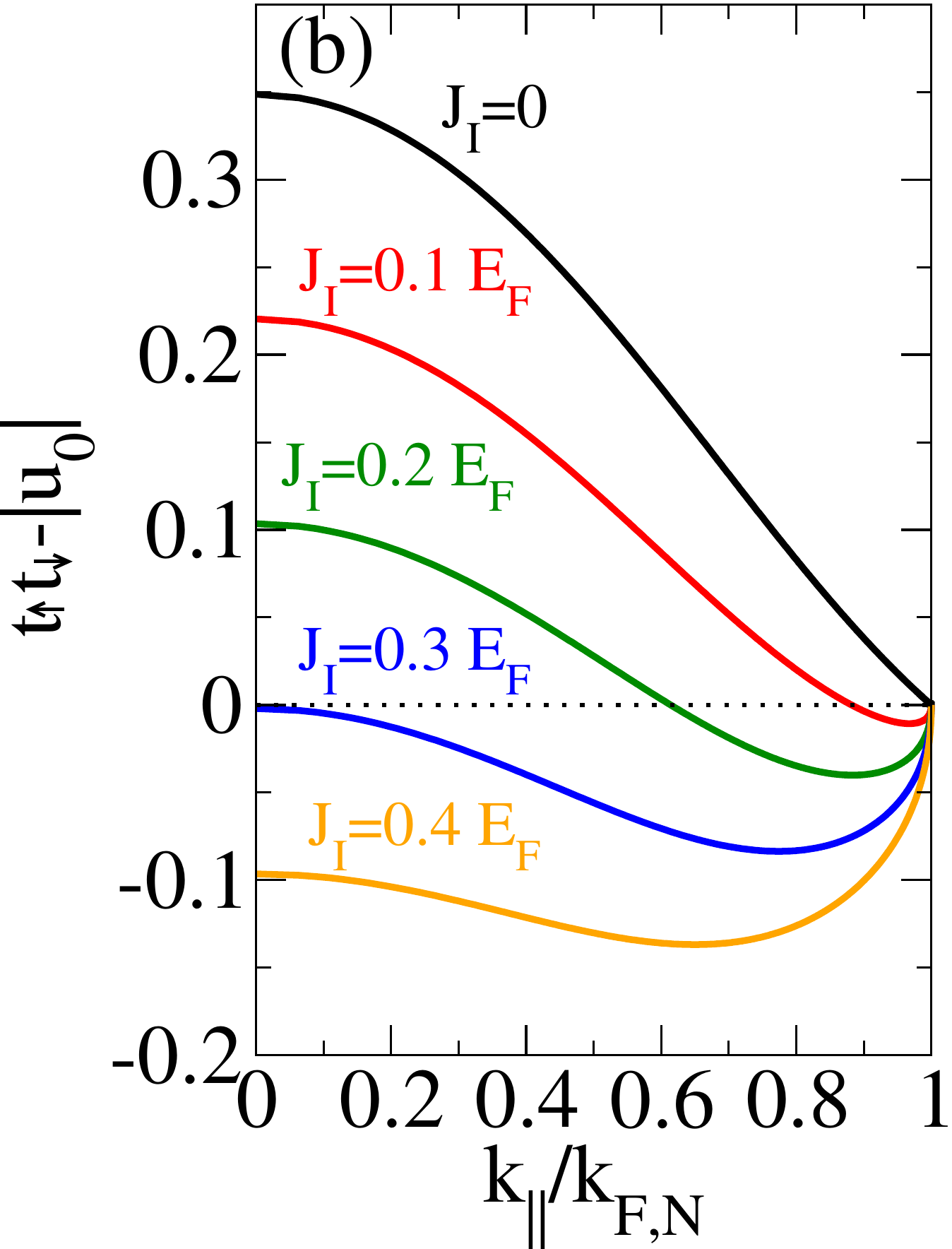}
\includegraphics[height=0.64\columnwidth]{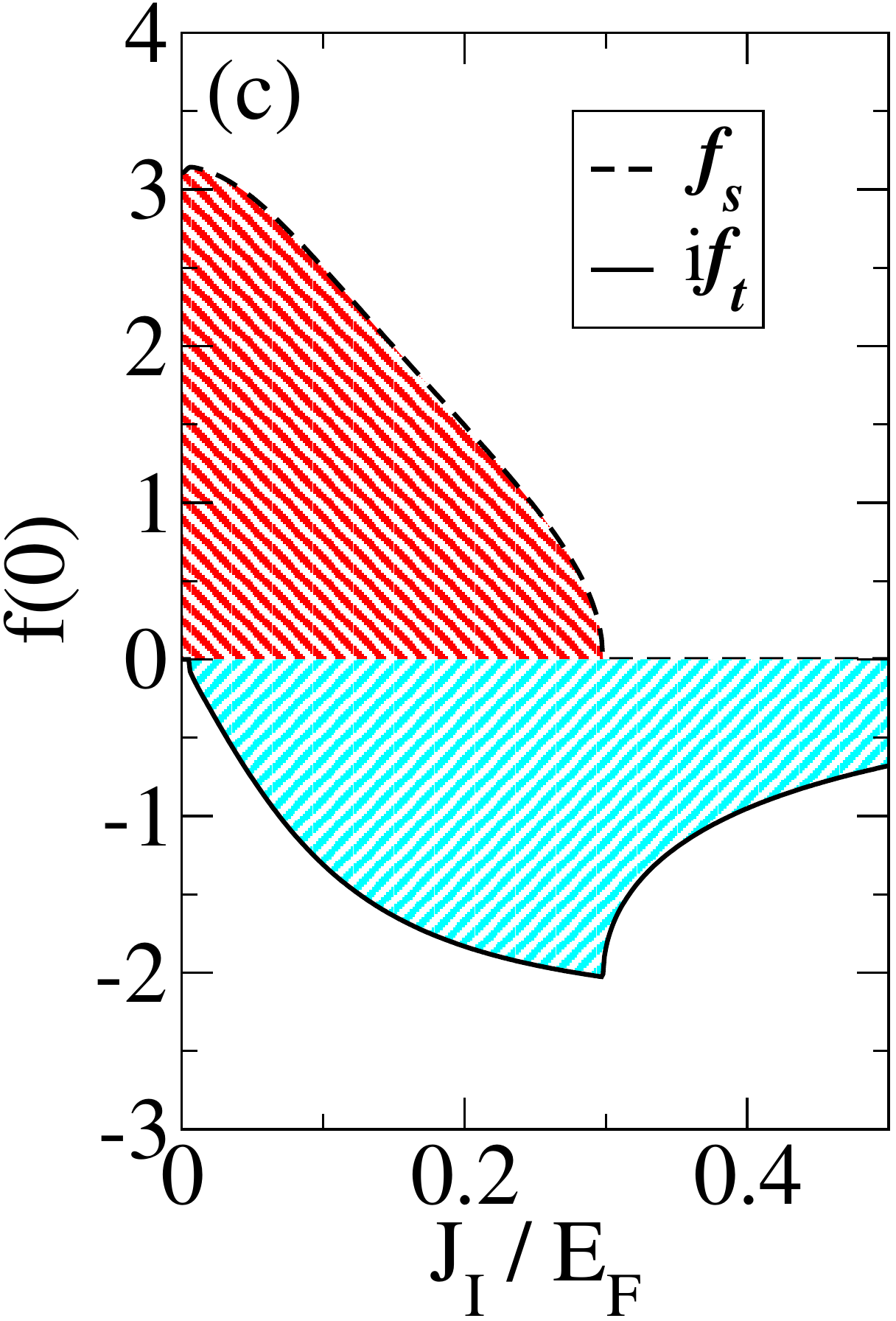}
\includegraphics[height=0.64\columnwidth]{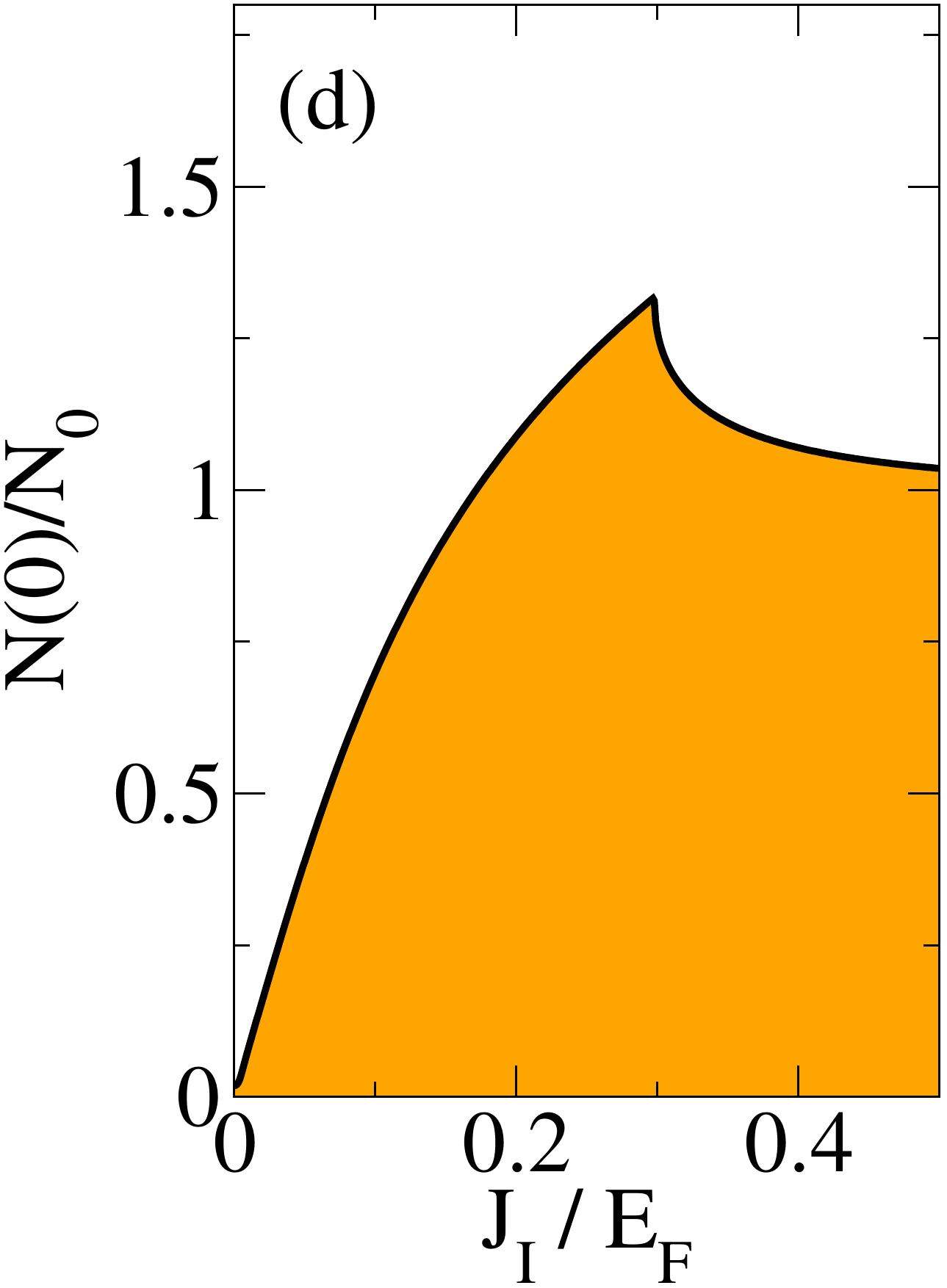}
\caption{(Color online) 
The same as in Fig.~\ref{fig:cleanDOS01}, however
with no Fermi surface mismatch, $k_{\rm F,N}=k_{\rm F,S}$.
\label{fig:cleanDOS1}}
\end{figure*}
\begin{figure*}
\includegraphics[height=0.63\columnwidth]{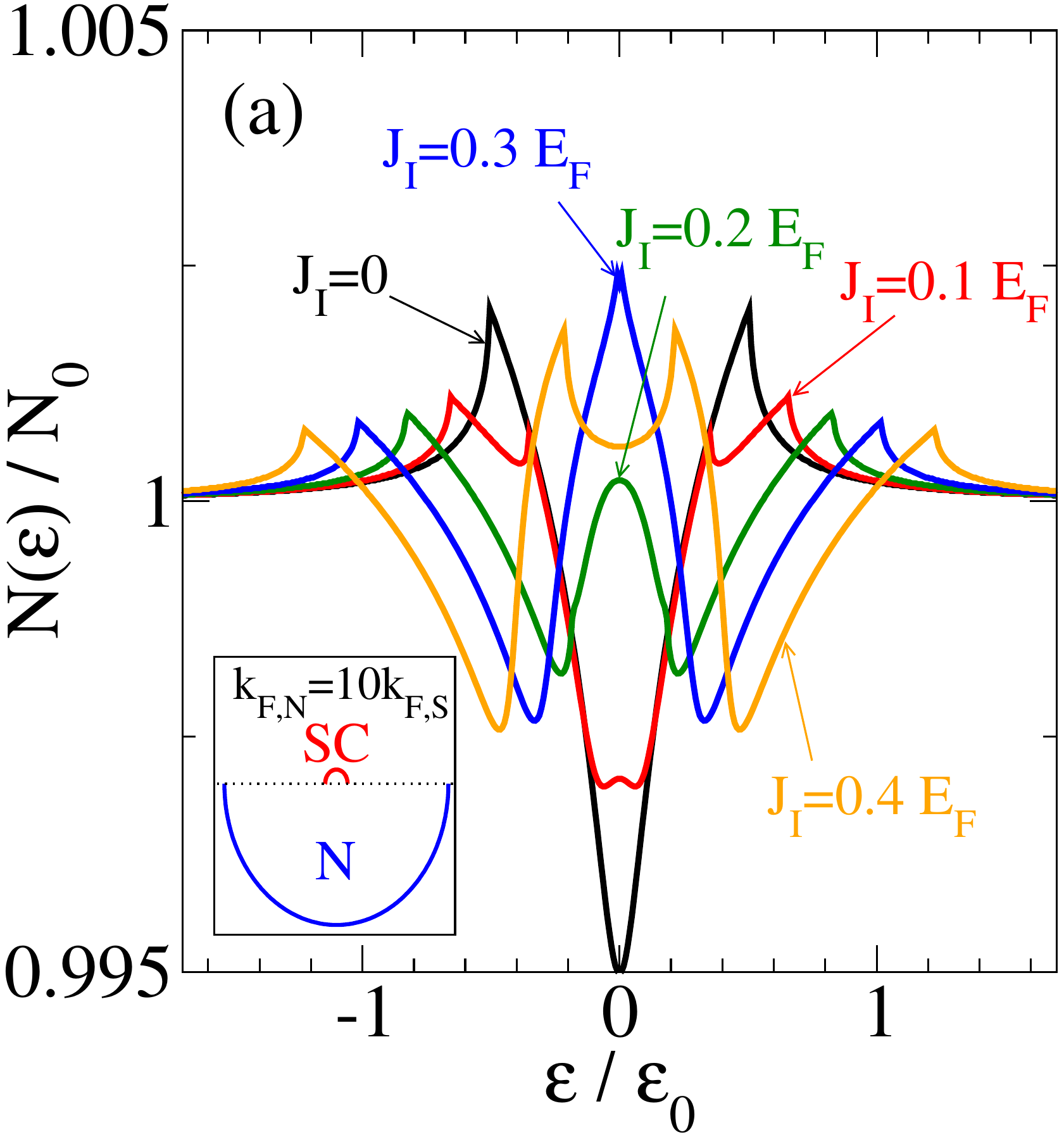}
\includegraphics[height=0.63\columnwidth]{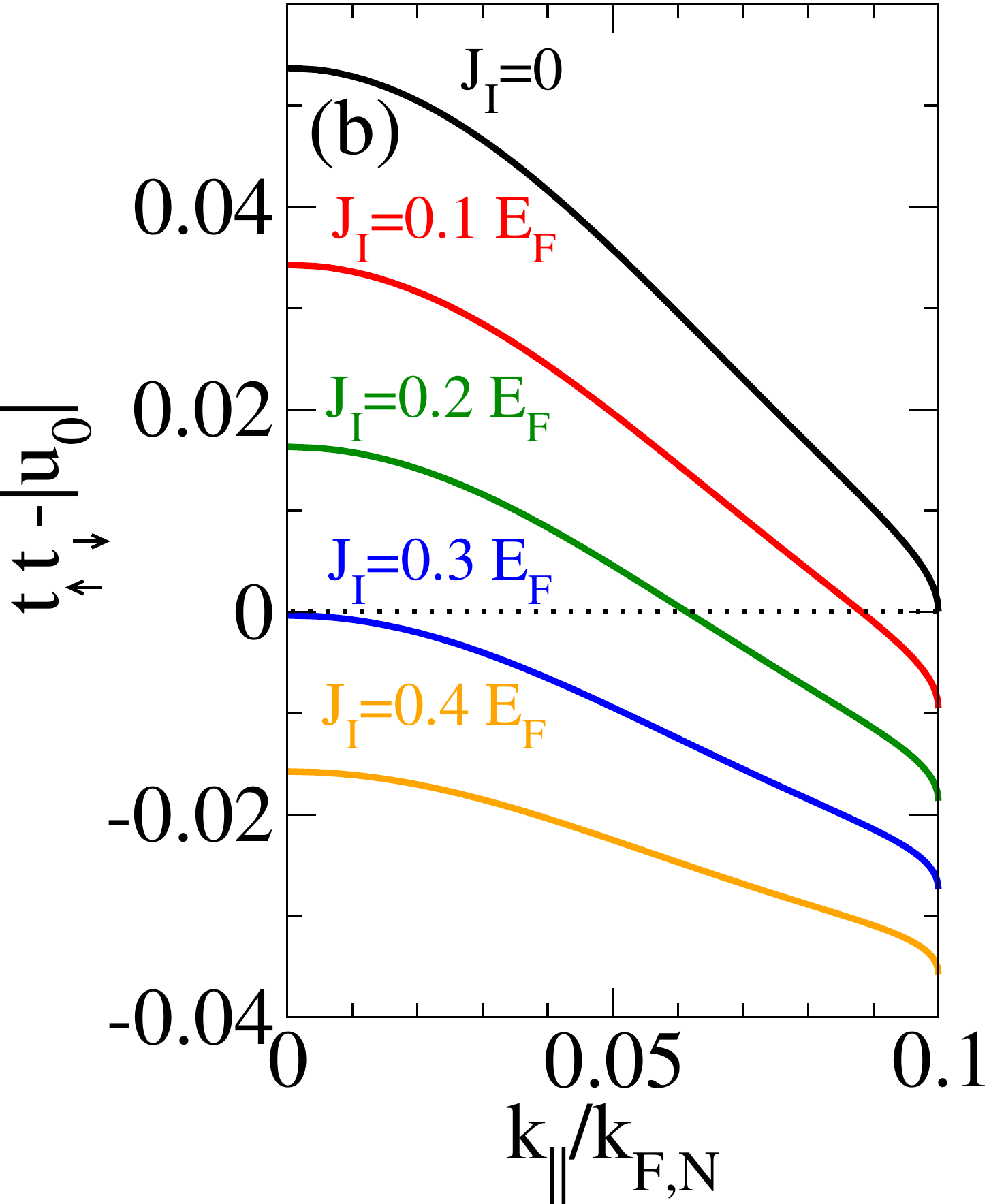}
\includegraphics[height=0.63\columnwidth]{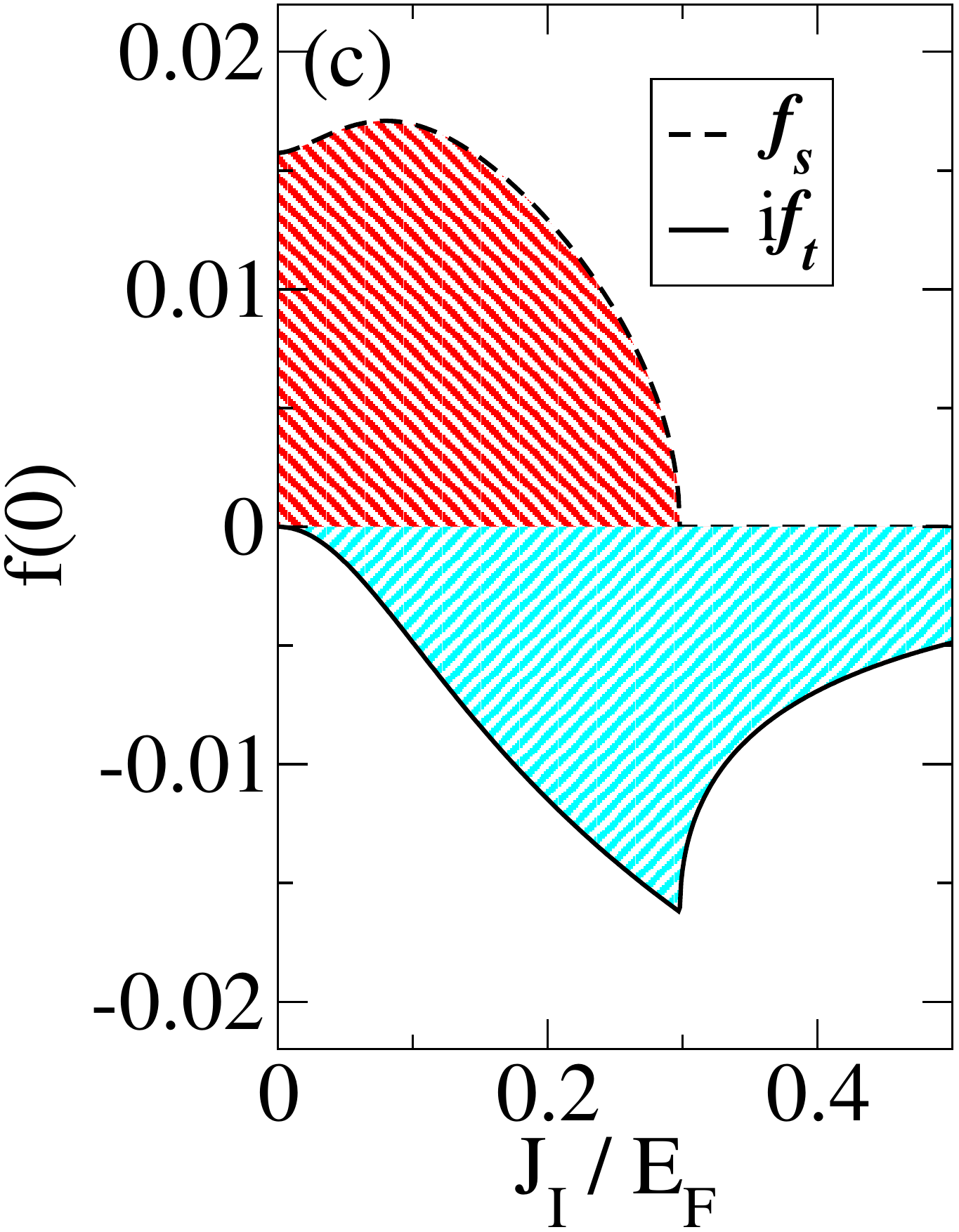}
\includegraphics[height=0.63\columnwidth]{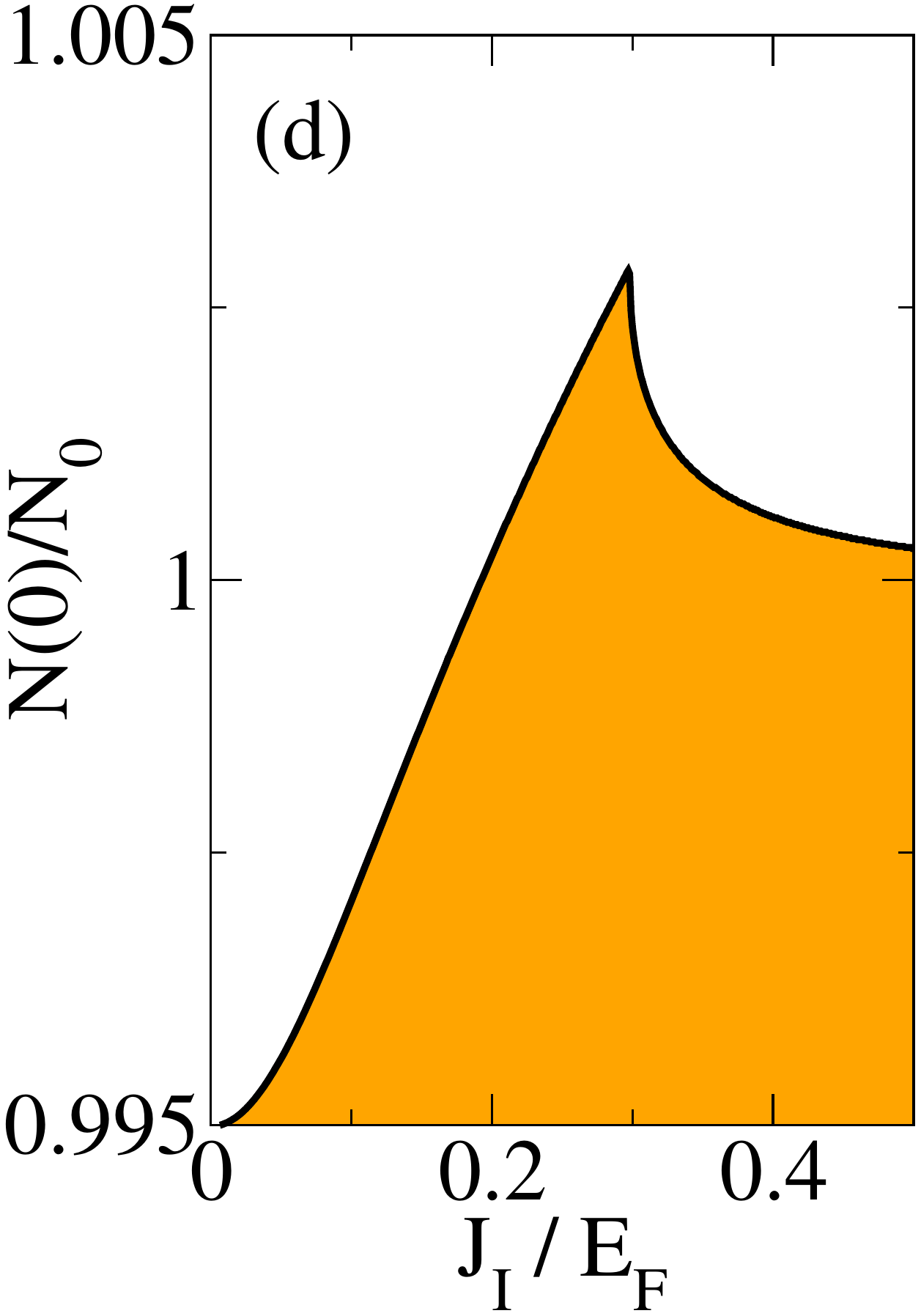}
\caption{(Color online) 
The same as in Fig.~\ref{fig:cleanDOS01}, however
with a different Fermi surface mismatch, $k_{\rm F,N}=10 k_{\rm F,S}$.
\label{fig:cleanDOS10}}
\end{figure*}
For all results presented below, the lower of the two potential barriers
is $V_\text{I}=0.2 E_\text{F}$, 
and the interface width is $d_\text{I}=2\lambda_{\text{F},\text{S}}/2\pi$.
We assume for simplicity equal band masses all over the system, and isotropic
Fermi surfaces. Thus, the energy dispersions are in the superconductor
$\ve{k}^2/2m$, in the normal metal $E_\text{N}+\ve{k}^2/2m $, and
in the barrier $E_\text{F}+U_{\text{I}\uparrow,\downarrow}+\ve{k}^2/2m $,
where $E_\text{F}=\ve{k}_{\text{F},\text{S}}^2/2m$. The constant
$E_\text{N}$ determines the Fermi surface mismatch between the superconductor
and the normal metal, with Fermi wavevectors $\ve{k}_{\text{F},\text{S}}$ and
$\ve{k}_{\text{F},\text{N}}$ for the superconductor and the normal metal,
respectively.
For such a model, the parameters in Eq.~\eqref{Rparameters} are given 
by
\begin{eqnarray}
\label{rpar}
\underline{R}_{\rm S}&=& \frac{\underline{\rho}_{\rm S}-\underline{v}^2 \underline{\rho}_{\rm N}}{\underline{1}-\underline{v}^2\underline{\rho}_{\rm N}\underline{\rho}_{\rm S}} , \quad
\underline{R}_{\rm N}= \frac{\underline{\rho}_{\rm N}-\underline{v}^2 \underline{\rho}_{\rm S}}{\underline{1}-\underline{v}^2\underline{\rho}_{\rm N}\underline{\rho}_{\rm S}} ,\\
\underline{T}_{\rm SN}&=& \frac{\underline{v}\; \, [(\underline{1}-\underline{\rho}_{\rm S}^2)(\underline{1}-\underline{\rho}_{\rm N}^2)]^{1/2}}{\underline{1}-\underline{v}^2\underline{\rho}_{\rm N}\underline{\rho}_{\rm S}}= \underline{T}_{\rm NS} ,
\label{tpar}
\end{eqnarray}
where $\underline{\rho}_j$ with $j \in \left\{ {\rm S},{\rm N}\right\}$, 
and $\underline{v}$ are diagonal spin-matrices 
with $\rho_{j,\sigma\sigma}=(k_{j}-\i\kappa_\sigma )/(k_{j}+\i\kappa_\sigma)$, and $v_{\sigma \sigma}=\exp(-\kappa_\sigma d_{\rm I})$.
Here, $\kappa_\sigma = [k_{\rm F,S}^2 U_{\rm I\sigma }/E_{\rm F}+k_{||}^2]^{1/2}$, and $k_{j}= [k_{{\rm F},j}^2-k_{||}^2]^{1/2}$.

We now turn to the discussion of our results,
shown in Figs.~\ref{fig:cleanDOS01}-\ref{fig:cleanDOS10}. 
The calculations were obtained for various Fermi surface geometries.
In Fig.~\ref{fig:cleanDOS01} we present results for the case
$k_{\rm F,S}\gg k_{\rm F,N}$, i.e. when the Fermi surface
mismatch is large, and the Fermi surface in the superconductor is much larger than that in the normal metal [see inset in Fig.~\ref{fig:cleanDOS01}(a)].
A more moderate mismatch is assumed in
Fig.~\ref{fig:cleanDOS05}, with $k_{\rm F,N}= 0.5 k_{\rm F,S}$.
In Fig.~\ref{fig:cleanDOS1} we consider the special case of no Fermi surface
mismatch, i.e.  $k_{\rm F,S}= k_{\rm F,N}$. Finally,
in Fig.~\ref{fig:cleanDOS10} we consider the case opposite to 
Fig.~\ref{fig:cleanDOS01}, namely a strong Fermi surface mismatch where the
Fermi surface in the normal metal is much larger than that in the superconductor,
$k_{\rm F,S}\ll k_{\rm F,N}$.
In each figure, we present in (a) the energy
resolved DOS for several spin polarizations of the interface barrier.
The energy scale of interest is the Thouless energy of the normal metal
layer, $\varepsilon_{\rm Th}=\hbar v_{\rm F,N}/2d_{\rm N}$, times the pair transmission amplitude from the superconductor to the
normal metal, $t_\uparrow t_\downarrow$. This quantity depends on the impact angle $\theta $; for definiteness we use as energy scale $\varepsilon_0 =t_\uparrow t_\downarrow \varepsilon_{\rm Th}$ for normal impact.
In (b) we show the quantity $t_\uparrow t_\downarrow-|u_0|$, 
where $u_0=u_{+}(\varepsilon=0)=-u_{-}(\varepsilon=0)$ is defined in Eq.~\eqref{u0}.
The plotted quantity controls the transition from even-frequency singlet 
to odd-frequency triplet correlations at the chemical potential ($\varepsilon=0$), 
as will be shown below.
The parameters $\vartheta_\text{S,N}$ and $r_{\uparrow,\downarrow}$ depend
on $k_{||}$, the momentum component parallel to the interface, which is conserved in the scattering process. Consequently, the parameter 
$t_\uparrow t_\downarrow-|u_0|$ depends on $k_{||}$
as well, and we show in the figure this dependence.
In (c) we show the even-frequency singlet and 
the odd-frequency triplet superconducting amplitudes at the chemical potential
[we plot the real quantities
$f_{\rm s}(\varepsilon)=0$ and $\i f_{\rm t}(\varepsilon=0)$)]
at the outer surface of the normal metal.
Finally, in (d) we show the local DOS at the 
chemical potential normalized to the normal state DOS, $N(\varepsilon=0)/N_0$, 
again at the outer surface of the normal metal.

We proceed with the discussion of the results.
We recall first the known behavior for zero interface spin polarization, $J_{\rm I}=0$ in
Figs.~\ref{fig:cleanDOS01}-\ref{fig:cleanDOS10}.
When all trajectories in the normal metal are partially transmissive
(Figs.~\ref{fig:cleanDOS01}-\ref{fig:cleanDOS1}), the DOS is zero at the
chemical potential, $\varepsilon=0$, and
shows an increase to finite values as function of energy $\varepsilon $. 
This increase is directly associated with the behavior of the topmost 
($J_{\rm I}=0$) curves in 
Figs.~\ref{fig:cleanDOS01}-\ref{fig:cleanDOS1}(b) for 
glancing impact, $k_{||}/k_{\rm F,N}\approx 1$.
When there are non-transmissive trajectories present in the normal layer (Fig.~\ref{fig:cleanDOS10}), the DOS at the chemical potential is finite.
In this case, as $k_{\rm F,S}<k_{\rm F,N}$, there is a background DOS resulting from the
non-transmissive directions, $k_{||}>k_{\rm F,S}$, in the normal metal; 
this background contribution is not associated with any superconducting pair correlations, 
and is nearly constant in energy and nearly temperature independent 
(considering typical superconducting energy scales).
All changes of the DOS related to superconductivity take place
on top of that background contribution. 

When the spin polarization of the interface increases to non-zero values, 
we can define three characteristically different regions of interface spin polarization 
$J_{\rm I}$. 
We turn our attention to panels (b) of
Figs.~\ref{fig:cleanDOS01}-\ref{fig:cleanDOS10}, which show the
quantity $t_\uparrow t_\downarrow -|u_0|$ as a function of $k_{||}$.
For directions where this quantity is positive, according to
Eq.~\eqref{eq:singlet} pure singlet correlations are 
created at the chemical potential in the normal metal, whereas for directions where this quantity is
negative, according to Eq.~\eqref{eq:triplet} pure odd-frequency triplet correlations are created at the chemical potential in the normal metal.
We can classify the curves into three groups, depending on the value of
$J_{\rm I}$.
We first have a region where 
$t_\uparrow t_\downarrow -|u_0|$ 
is positive for all $k_{||}$ (region I; e.g. $J_{\rm I}=0.2 E_{\rm F}$ in Fig.~\ref{fig:cleanDOS01}); second a region where
$t_\uparrow t_\downarrow -|u_0|$ 
is positive for some, and negative for other values of $k_{||}$ (region II; e.g. $J_{\rm I}=0.2 E_{\rm F}$ in Fig.~\ref{fig:cleanDOS1}); 
and third a region where 
$t_\uparrow t_\downarrow -|u_0|$ 
is negative for all $k_{||}$ (region III; e.g. $J_{\rm I}>0.3 E_{\rm F}$ in all four figures).

In Figs.~\ref{fig:cleanDOS01}-\ref{fig:cleanDOS10}(c) we show the singlet ($f_{\rm s}$) and triplet ($f_{\rm t}$) component 
of the momentum-averaged, i.e. $s$-wave, correlation functions at the chemical potential. In general, there also exist 
higher order even-parity components which behave qualitatively similar.
When increasing $J_\text{I}$ in region I, it can be seen from Fig.~\ref{fig:cleanDOS01} and
\ref{fig:cleanDOS05} that the pair correlations at the chemical potential stay purely singlet,
and the DOS at the Fermi level, shown in (d), stays zero. 
When $J_\text{I}$ enters region II, there is a strong mixing between singlet and triplet amplitudes, 
and the DOS at the Fermi level rises to non-zero values. 
Finally, when $J_\text{I}$ is above $J_\text{crit}$ (region III) 
the singlet correlations vanish identically at the chemical potential, and pure odd-frequency triplet 
amplitudes remain, when the DOS is larger than its normal state value. The transition from the region 
III can be identified as a sharp decrease of the DOS as function of $J_\text{I}$ from a maximum value
in Figs.~\ref{fig:cleanDOS01}-\ref{fig:cleanDOS10}. 

Region I only exists for $k_{\rm F,N} < k_{\rm F,S}$ (Figs.~\ref{fig:cleanDOS01} and \ref{fig:cleanDOS05}).
As seen from Fig.~\ref{fig:cleanDOS01}(d),
as long as $k_{\rm F,N}\ll k_{\rm F,S}$ the DOS shows 
just as in the diffusive case 
a rather sharp transition from $N(0)=0$ 
to a value above the normal-state DOS
as a function of the interface exchange splitting $J_{\rm I}$.
The existence of region II in
Figs.~\ref{fig:cleanDOS05}-\ref{fig:cleanDOS10}
is due to the fact that
the mixing angle drops slower with impact angle than the transmission. 
It is characterized by a zero crossing of the parameter
$t_\uparrow t_\downarrow-|u_0|$ as function of parallel momentum $k_{||}$.
For increasing $k_{\rm F,N}$ region II
extends to lower values of $J_{\rm I}$, and
when $k_{\rm F,N} \ge k_{\rm F,S}$, region II starts at $J_{\rm I}=0$ and extends to
a critical value $J_{\rm crit}$. This is due to the fact that for any small $J_{\rm I}\ne 0$
there are negative values of $t_\uparrow t_\downarrow -|u_0|$ for the 
largest transmissive $k_{||}$. For $k_{\rm F,N}>k_{\rm F,S}$ this 
can be understood easily because
the transmission probability goes to zero whereas the
spin-mixing angles stay finite when $k_{||}$ approaches $k_{\rm F,S}$.
For $J_{\rm I}>J_{\rm crit}$ the system is in region III.
For any mismatch between the Fermi surfaces, there is a critical value $J_{\rm crit}$.

\begin{figure}[b]
\includegraphics[width=0.99\columnwidth]{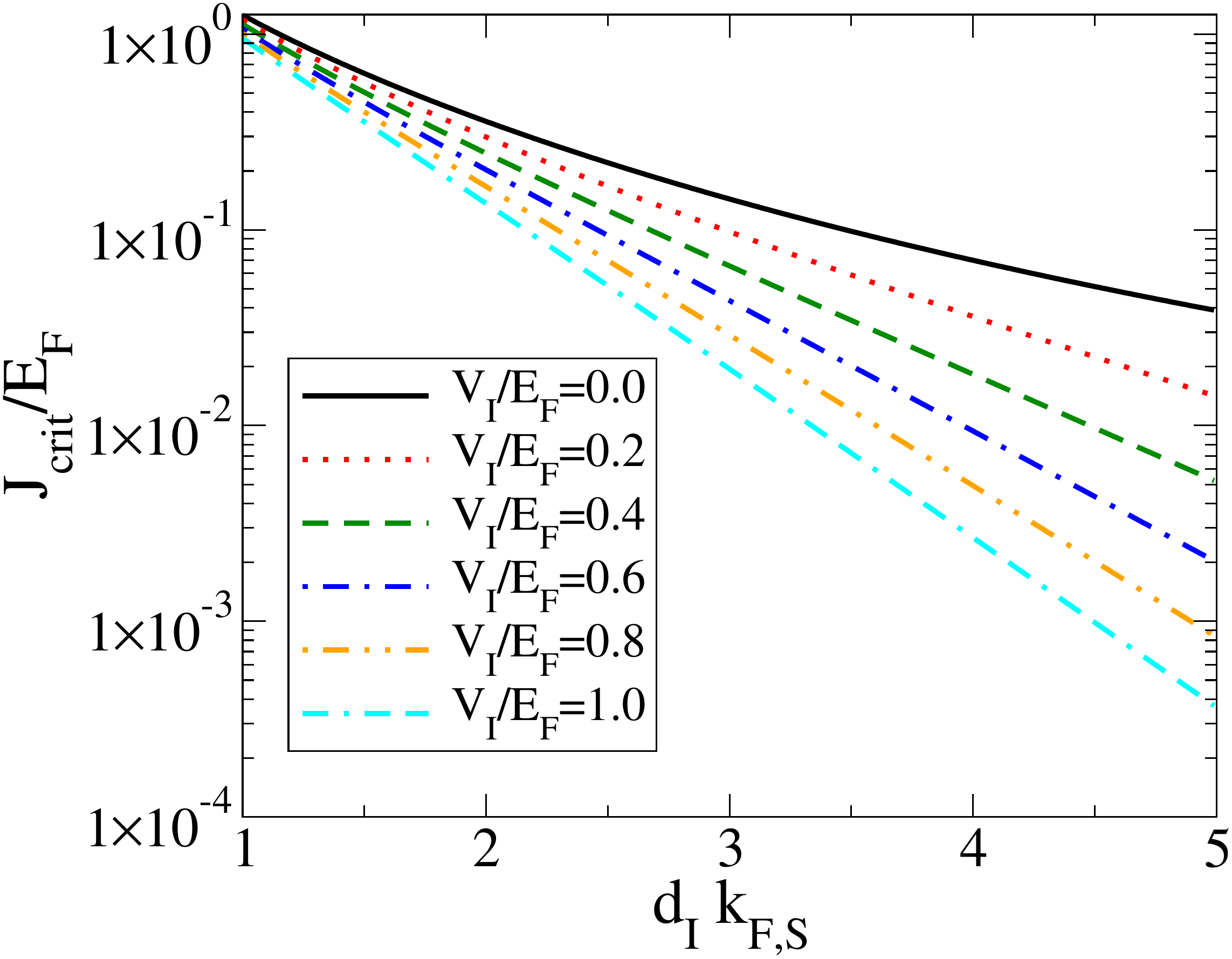}
\caption{(Color online) 
The critical value $J_{\rm crit}$ as function of interface thickness for
various strengths of the interface potential. The curves were obtained by
finding numerically the solution of Eq.~\eqref{jcrit}.}
\label{fig:Jcrit}
\end{figure}
It is interesting to note that,
although both the spin-mixing angles and the transmission
probabilities vary with Fermi surface mismatch,
for a box-shaped potential the critical value $J_{\rm crit}$
does not depend on the ratio $k_{\rm F,N}/k_{\rm F,S}$. Thus, it has the same value,
$J_{\rm crit} \approx 0.3 E_{\rm F}$, in Figs.~\ref{fig:cleanDOS01}-\ref{fig:cleanDOS10}. 
This value is determined by 
the condition that $t_\uparrow t_\downarrow=|u_0|$ for $k_{||}=0$.
Inserting Eqs.~\eqref{rpar} and \eqref{tpar} into Eq.~\eqref{Rparameters},
and using Eqs. \eqref{eq:smatrix} and \eqref{u0},
this condition leads to the following implicit equation for
the value of $J_{\rm crit}$
\begin{eqnarray}
4&=&
\left(1+\frac{1}{\nu_{\uparrow} \nu_{\downarrow } }\right)
\left(\nu_{\downarrow }-\nu_{\uparrow }\right)
\sinh\left[(\nu_{\downarrow }+\nu_{\uparrow })\delta_{\rm I}\right]\nonumber \\
&+&
\left(1-\frac{1}{\nu_{\uparrow }\nu_{\downarrow }}\right)
\left(\nu_{\downarrow }+\nu_{\uparrow }\right)
\sinh\left[(\nu_{\downarrow }-\nu_{\uparrow })\delta_{\rm I}\right] , \quad
\label{jcrit}
\end{eqnarray}
with the parameters $\nu_{\downarrow }=\sqrt{(V_{\rm I}+J_{\rm crit}) /E_{\rm F}}$,
$\nu_{\uparrow }=\sqrt{V_{\rm I}/E_{\rm F}}$, and 
$\delta_{\rm I}= k_{\rm F,S} \, d_{\rm I} $.
Solutions of this equation are shown in Fig.~\ref{fig:Jcrit}.
We find that the transition occurs earlier for thicker interfaces. 
This is because the transmission decreases with interface width, while the mixing 
angle is actually enhanced to some extent, as discussed in Ref.~\onlinecite{greinPRB}.
In order to both achieve a satisfying transmission, and to have realistic
values for the exchange field, $d_{\rm I} k_{\rm F,S}$ should
be between 1 and 2.
The remarkable robustness of the critical interface spin-polarization with respect to the
Fermi-surface mismatch might simplify the experimental task to observe this effect, as
the usual restrictions for finding suitable materials to match at the interface are relaxed. 
We caution, however, that the above strict independence on the Fermi-surface mismatch might
be relaxed for more realistic interface potentials.

\section{Summary}\label{sec:summary}

In this work, we have provided a comprehensive treatment of the proximity effect in a system consisting of a normal metal (e.g. Cu)
in contact with a conventional $s$-wave superconductor (e.g. Al)
through a spin-active interface. Such a spin-active interface is 
incorporated by using, e.g., a ferromagnetic insulator such as EuO. We have shown that based on the self-consistent 
calculation in the diffusive regime, the even-odd frequency conversion first predicted in Ref.~\onlinecite{linder_prl_09} 
is robust even when taking into account pair-breaking effects near the interface which cause a depletion of the superconducting 
order parameter. Although the conversion relies crucially on interface properties which vary considerably with the impact angle 
of incident quasiparticles, it is generically robust against Fermi-surface averaging in the clean limit. Moreover, we show 
that the conversion takes place even when the superconducting region does not act as a reservoir, i.e. when the thicknesses 
of the superconducting and normal layers are comparable. Our findings suggest a robust and simple method of obtaining a  
clear-cut experimental signature of odd-frequency superconducting correlations. 

\acknowledgments

We would like to thank W. Belzig, G. Sch\"on, and E. Zhao
for helpful contributions. J.L. and A.S. were supported by the 
Norwegian Research Council Grant No. 167498/V30 (STORFORSK).


\begin{thebibliography}{99}

\bibitem{bergeretrmp} F. S. Bergeret, A. F. Volkov, and K. B. Efetov, 
Rev. Mod. Phys. \textbf{77}, 1321 (2005).

\bibitem{buzdinrmp} A. I. Buzdin, Rev. Mod. Phys. \textbf{77}, 935 (2005).

\bibitem{eschrig07}
M. Eschrig, T. L\"ofwander, T. Champel, J. C. Cuevas, J. Kopu, and G. Sch\"on,
J. Low. Temp. Phys. {\bf 147}, 457 (2007).

\bibitem{bergeretPRL} F. S. Bergeret, A. F. Volkov, and K. B. Efetov, Phys. Rev. Lett. \textbf{86}, 4096 (2001).

\bibitem{berezinskii} V. L. Berezinskii, 
Pis'ma Zh. Eksp. Teor. Fiz. \textbf{20}, 628 (1974);
[JETP Lett. {\bf 20}, 287 (1974)].

\bibitem{bal92}
A. Balatsky and E. Abrahams,
Phys. Rev. B {\bf 45}, 13125 (1992).

\bibitem{abr95}
E. Abrahams, A. Balatsky, D.~J. Scalapino, and J.~R. Schrieffer,
Phys. Rev. B {\bf 52}, 1271 (1995).

\bibitem{coleman93}
P. Coleman, E. Miranda, and A. Tsvelik, Phys. Rev. Lett. {\bf 70}, 2960 (1993).

\bibitem{fuseya03}
Y. Fuseya, H. Kohno, and K. Miyake, Journ. Phys. Soc. Jap. {\bf 72}, 2914 (2003).

\bibitem{tanakaPRL} Y. Tanaka, A. A. Golubov, S. Kashiwaya, and M. Ueda, Phys. Rev. Lett. \textbf{99}, 037005 (2007).

\bibitem{Yokoyama} T. Yokoyama, Y. Tanaka, and A. A. Golubov, Phys. Rev. B {\bf 78}, 012508 (2008).



\bibitem{Yokoyama2} T. Yokoyama, M. Ichioka, and Y. Tanaka, J . Phys. Soc. Jpn. {\bf 79} 034702 (2010).

\bibitem{yokoyama07} T. Yokoyama, Y. Tanaka, and A. A. Golubov, Phys. Rev. B \textbf{75}, 134510 (2007).

\bibitem{eschrig08} 
M. Eschrig and  T. L{\"o}fwander, Nature Physics {\bf 4}, 138 (2008).

\bibitem{linderyokoyama_prb_08} J. Linder, T. Yokoyama, and A. Sudb{\o}, Phys. Rev. B \textbf{77}, 174507 (2008).

\bibitem{linderzareyan_prb_09} J. Linder, M. Zareyan, and A. Sudb{\o}, Phys. Rev. B \textbf{80}, 014513 (2009).

\bibitem{eschrig03} M. Eschrig, J. Kopu, J. C. Cuevas, and Gerd Sch{\"o}n, Phys. Rev. Lett. \textbf{90}, 137003 (2003). 

\bibitem{keizer_nature_06}  R. S. Keizer, S. T. B. Goennenwein, T. M. Klapwijk, G. Miao, G. Xiao, and A. Gupta, Nature \textbf{439}, 825 (2006).

\bibitem{sosnin06}
I. Sosnin, H. Cho, V. T. Petrashov, and A. F. Volkov, Phys. Rev. Lett. {\bf 96}, 157002 (2006).

\bibitem{greinPRL} 
R. Grein, M. Eschrig, G. Metalidis, and G. Sch\"on, Phys. Rev. Lett. {\bf 102}, 227005 (2009).

\bibitem{greinPRB} 
R. Grein, T. L\"ofwander, G. Metalidis, and M. Eschrig, 
Phys. Rev. B {\bf 81}, 094508 (2010).

\bibitem{zaitsev84}
A.~V. Zaitsev, Zh. Eksp. Teor. Fiz. {\bf 86}, 1742 (1984)
[Sov. Phys. JETP {\bf 59}, 1015 (1984)].

\bibitem{shelankov84}
A. L. Shelankov, Fiz. Tverd. Tela (Leningrad) {\bf 26}, 1615 (1984) [Sov. Phys. Solid State {\bf 26}, 981 (1984)].

\bibitem{eschrig00} M. Eschrig, Phys. Rev. B \textbf{61}, 9061 (2000). 


\bibitem{kupluk} M. Yu. Kupriyanov and V. F. Lukichev, Zh. Exp. Teor. Fiz. \textbf{94}, 139 (1988) . 

\bibitem{nazarov99} 
Yu. Nazarov, Superlatt.Microstruct. \textbf{25}, 1221 (1999). 

\bibitem{meservey}
R. Meservey and P. M. Tedrow, Phys. Rep. {\bf 238}, 173 (1994).

\bibitem{Tokuyasu88}
T. Tokuyasu, J. A. Sauls, and D. Rainer, Phys. Rev. B {\bf 38}, 8823 (1988).

\bibitem{hh} D. Huertas-Hernando, Yu.V. Nazarov, W. Belzig, arXiv:cond-mat/0204116; D. Huertas-Hernando, Yu. V. Nazarov, and W. Belzig, Phys. Rev. Lett. \textbf{88}, 047003 (2002).

\bibitem{audrey} A. Cottet and W. Belzig, Phys. Rev. B \textbf{72}, 180503 (2005); A. Cottet and J. Linder, Phys. Rev. B \textbf{79}, 054518 (2009); A. Cottet, D. Huertas-Hernando, W. Belzig, and Y. V. Nazarov, Phys. Rev. B  \textbf{80}, 184511 (2009). 
.

\bibitem{bobkova07}
I. V. Bobkova and A. M. Bobkov, Phys. Rev. {\bf 76}, 094517 (2007).

\bibitem{linder_prb_07} J. Linder and A. Sudb{\o}, Phys. Rev. B \textbf{75}, 134509 (2007); J. Linder, T. Yokoyama, and A. Sudb{\o}, Phys. Rev. B \textbf{79}, 054523 (2009).

\bibitem{brydon} P. M. Brydon and D. Manske, Phys. Rev. Lett. \textbf{103}, 147001 (2009); P. M. Brydon, B. Kastening, D. K. Morr, and D. Manske, Phys. Rev. B \textbf{77}, 104504 (2008).

\bibitem{lindercuoco_arxiv_10} J. Linder, M. Cuoco, and A. Sudb{\o}, arXiv:1003.0893.

\bibitem{fogelstrom00} 
M. Fogelstr{\"o}m, Phys. Rev. B \textbf{62}, 11812 (2000). 

\bibitem{zhao04} 
E. Zhao, T. L{\"o}fwander, and J. A. Sauls, Phys. Rev. B \textbf{70}, 134510 (2004).

\bibitem{Millis} A. Millis, D. Rainer, and J. A. Sauls, Phys. Rev. B {\bf 38}, 4504 (1988). 
\bibitem{eschrig09}
M. Eschrig, Phys. Rev. B {\bf 80}, 134511 (2009).

\bibitem{linder_prl_09} J. Linder, T. Yokoyama, A. Sudb{\o}, and M. Eschrig, Phys. Rev. Lett. \textbf{102}, 107008 (2009).


\bibitem{qcl1} 
J. W. Serene and D. Rainer, Phys. Rep. \textbf{101}, 221 (1983). 
\bibitem{qcl2} 
A. I. Larkin and Y. N. Ovchinnikov, in {\it Nonequilibrium Superconductivity},
edited by D. N. Langenberg and A. I. Larkin (Elsevier Science Publishers,
1986), p. 493.
\bibitem{qcl3} 
A. Schmid and G. Sch\"on, J. Low. Temp. Phys. {\bf 20}, 207 (1975).
\bibitem{qcl4} 
N. Kopnin, \textit{Theory of Nonequilibrium Superconductivity},
(Oxford University Press, New York, 2001).
\bibitem{qcl5} 
M. Eschrig, J. A. Sauls, H. Burkhardt, and D. Rainer, in 
{\it High-T$_c$ Superconductors and Related Materials, Fundamental 
Properties, and Some Future Electronic Applications}, Proc.
NATO Advanced Study Institute, edited by S.-L. Drechsler and
T. Mishonov, pp. 413-446 (Kluwer Academic, Norwell, MA, 2001).

\bibitem{linder_prb_08} J. Linder, T. Yokoyama, and A. Sudb{\o}, Phys. Rev. B \textbf{77}, 174514 (2008).

\bibitem{bruder} C. Bruder, Phys. Rev. B \textbf{41}, 4017 (1990).

\bibitem{nagato}
Y. Nagato, K. Nagai, and J. Hara, J. Low Temp. Phys. {\bf 93}, 33 (1993); S. Higashitani and K. Nagai, J. Phys. Soc. Jpn. {\bf 64}, 549 (1995); Y. Nagato, S. Higashitani, K. Yamada, and K. Nagai, J. Low Temp. Phys. {\bf 103}, 1 (1996).

\bibitem{schopohl}
N. Schopohl and K. Maki, Phys. Rev. B {\bf 52}, 490 (1995); N. Schopohl, arXiv:cond-mat/9804064 (unpublished).

\bibitem{cuevas06}
J. C. Cuevas, J. Hammer, J. Kopu, J. K. Viljas, and M. Eschrig, Phys. Rev. B {\bf 73}, 184505 (2006).

\bibitem{usadel} K. Usadel, Phys. Rev. Lett. \textbf{25}, 507 (1970).

\bibitem{Asano} Y. Asano, Y. Tanaka, and A. A. Golubov, Phys. Rev. Lett. {\bf 98}, 107002 (2007). 

\bibitem{Braude} V. Braude and Yu. V. Nazarov, Phys. Rev. Lett. {\bf 98}, 077003 (2007).

\bibitem{clogston} A. M. Clogston, Phys. Rev. Lett. \textbf{9}, 266 (1962);
B. S. Chandrasekhar, Appl. Phys. Lett. \textbf{1}, 7 (1962).

\end{thebibliography}
\end{document}